\documentclass[pre, aps,twocolumn]{revtex4}
\newcommand{\EQ}[1]{Eq.~(\ref{eq:#1})}
\newcommand{\EQS}[2]{Eqs.~(\ref{eq:#1}) and (\ref{eq:#2})}
\newcommand{\FIG}[1]{Fig.~\ref{fig:#1}}

\usepackage{verbatim}
\usepackage{hyperref}

\usepackage{graphicx}
\usepackage{amsmath}

\newcommand{\suppfigSFS}{8}
\newcommand{\suppfigTMRCA}{7}
\newcommand{\suppfigLargeN}{6}

\newcommand{\avgmut}{\Delta_\mu}
\newcommand{\x}{x}
\newcommand{\xu}{y}
\newcommand{\xs}{x}
\newcommand{\Ne}{\tilde{N}}
\newcommand{\mfit}{\bar{\xu}}
\newcommand{\OL}{\mathcal{L}}

\newcommand{\Ai}{\mathrm{Ai}}
\newcommand{\affigure}{4}
\newcommand{\EQbsrates}{5}
\newcommand{\figAncestors}{Fig.~5A}

\begin{document}

\title{Genealogies of rapidly adapting populations}
\date{\today} 
\author{Richard A.~Neher}
\affiliation{Evolutionary Dynamics and Biophysics Group, Max Planck Institute for Developmental Biology, 72076, T\"ubingen, Germany}
\author{Oskar Hallatschek}
\affiliation{Biophysics and Evolutionary Dynamics Group, Max Planck Institute for Dynamics and Self-Organization, 37077 G\"{o}ttingen, Germany}

\begin{abstract}
The genetic diversity of a species is shaped by its recent evolutionary history
and can be used to infer demographic events or selective sweeps. Most inference
methods are based on the null hypothesis that natural selection is a weak or
infrequent evolutionary force.  However, many species, particularly pathogens,
are under continuous pressure to adapt in response to changing environments.  A
statistical framework for inference from diversity data of such populations is
currently lacking. Toward this goal, we explore the properties of genealogies
in a model of continual adaptation in asexual populations.  
We show that lineages trace back
to a small pool of highly fit ancestors, in which almost simultaneous
coalescence of more than two lineages frequently occurs. While such multiple
mergers are unlikely under the neutral coalescent, they create a unique genetic
footprint in adapting populations. The site frequency spectrum of derived
neutral alleles, for example, is non-monotonic and has a peak at high
frequencies, whereas Tajima's $D$ becomes more and more negative with increasing
sample size.
Since multiple merger coalescents emerge in many models of rapid adaptation, we
argue that they should be considered as a null-model for adapting populations.
\end{abstract}

\maketitle

Evolutionary change is usually too slow to be observed in real time. A
sequence sample represents a static snapshot from which we want to
learn about a dynamic evolutionary process. 
The predominant framework to analyze such population genetic data and infer
demographic history is Kingman's neutral coalescent. Within this model,
all individuals are equivalent, i.e., there are no fitness differences, and pairs of lineages merge at random. The
statistical properties of genealogies in this simple population
genetic model can be computed
exactly~\cite{Kingman:1982p28911,Derrida:1991p1707}, facilitating
comparison to data.
One central prediction of the neutral coalescent is that the genetic diversity
of a population is proportional to its size. This prediction, however, is at odds
with the observed weak correlation between genetic diversity and population
size, a paradox often remedied by the definition of an effective population size via
the genetic diversity. The model has been generalized to account for historic changes in
population size, mutation rates, geographical structure and the effects of
purifying selection
\cite{Nordborg:1997p43854,Charlesworth:1993p36005,Walczak:2011p45228,Ofallon:2010p43263,Barton:2004p34826}.
Positive selection, however, has proved difficult to incorporate, and progress
has been limited to rare selective sweeps
\cite{Barton:1998p28270,Durrett:2005p40919} and weak selection
\cite{Krone:1997p19666}.

\begin{figure}[btp]
\begin{center}
  \includegraphics[width=\columnwidth]{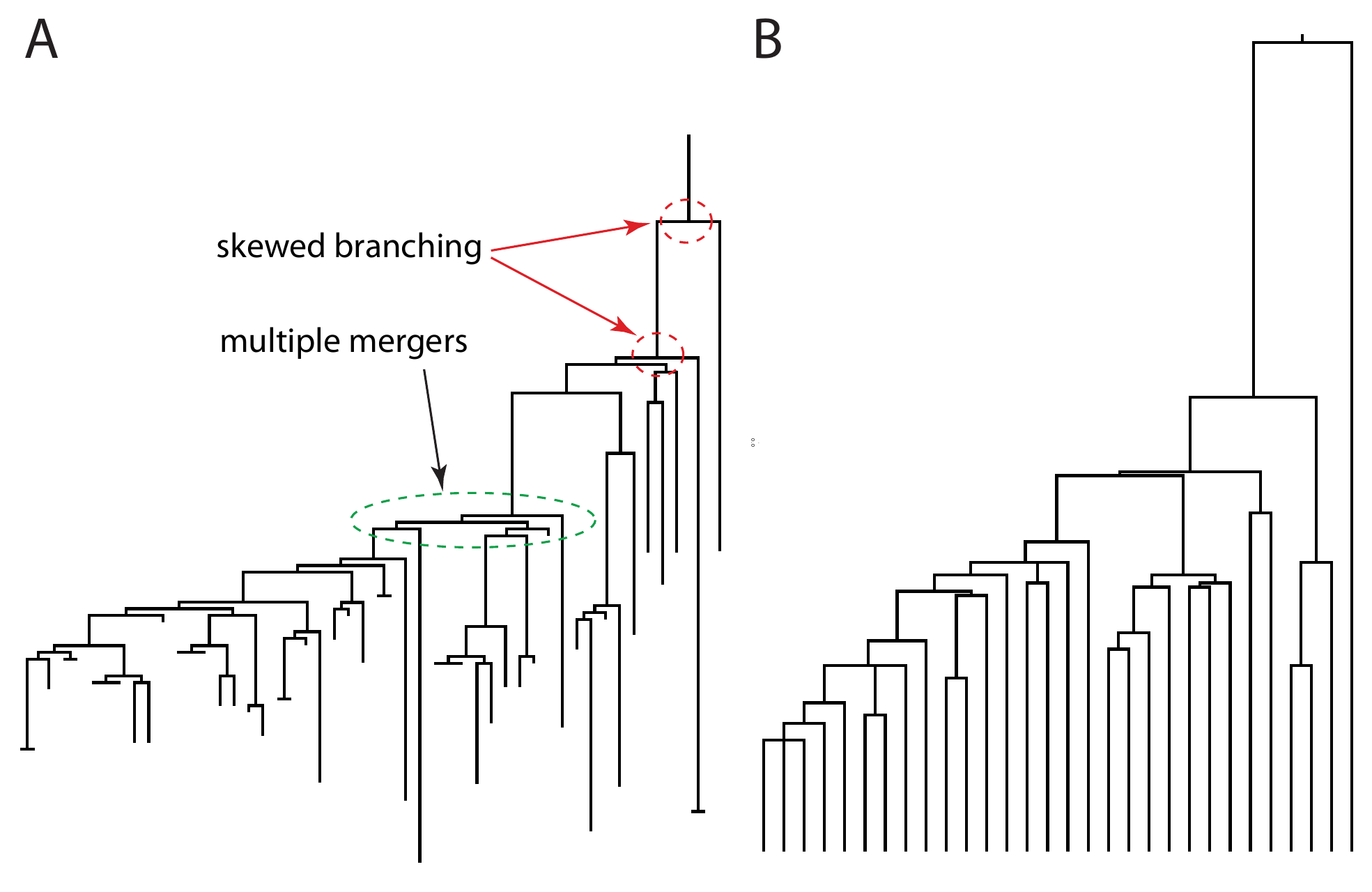}
  \caption{
  Panel A shows a maximum-likelihood tree of influenza nucleotide sequences (HA segment) sampled in Asia in 2009 (subtype H3N2) produced using Fasttree \cite{Price:2009p47657}.
  Panel B shows a tree drawn from a simulation of our model of adapting populations.   \label{fig:influenza}
  Both trees often branch very unevenly with almost all descendants on the left-most branch. While approximate multiple mergers are common in both trees, the influenza tree does not display the uniformly long terminal branches
  we observe in simulations. This could be due to heterogeneous sampling of influenza. Trees are drawn with Figtree \url{http://tree.bio.ed.ac.uk/software/figtree/}.}
\end{center}
\end{figure}

In many populations, particularly large microbial populations, selection is
neither rare nor weak. Instead, these populations are under sustained pressure to
adapt to changing environments. Prominent examples include pathogens like
influenza that continuously evade human immune responses
 or HIV, which establishes a chronic
infection despite heavy immune predation.
The genealogical trees reconstructed from sequence samples often suggest
substantial departure from neutrality; see
\cite{Bedford:2011p47600} for examples from viral evolution or
\cite{Seger:2010p39561} for eukaryotic examples. The influenza tree shown in
\FIG{influenza}, for instance, is incompatible with a neutral genealogy, since
there are parts where many lineages merge in a very brief period, and the tree
often branches extremely unevenly with very few individuals on one branch and
many on the other.
These two observations represent fundamental deviations from the standard
neutral model, even when a varying population size is allowed. Strelkowa and L\"assig present 
a detailed analysis of Influenza A evolution and conclude that influenza is governed by coalescence processes different from the Kingman coalescent \cite{Strelkowa:2012p48466} .

To analyze and interpret genealogies of populations under sustained
directional selection, an alternative simple null model would be
extremely useful. The features of genealogies discussed above are in
fact common to a class of non-Kingman coalescence models, which have
received considerable attention in the mathematical coalescent
literature \cite{Pitman:1999p41045,Berestycki:2009p45808}. A special
case is the Bolthausen-Sznitman
coalescent (BSC) \cite{Bolthausen:1998p47390}, which has been shown to
describe the genealogies in models where a population expands into
uninhabited territory \cite{Brunet:2007p18866}.  On the basis of a
particular exact solution and a phenomenological theory, Brunet {\it et
  al.}~conjectured that genealogies in all models of the same universality class (the class of 
stochastic Fisher-Kolmogorov-Petrovskii-Piscounov (FKPP) waves \cite{Fisher:1937p47380,Kolmogorov:1937p47381})
are described by the BSC; see \cite{Brunet:2012p45985} for a recent review. 
This universality class contains all models with short range 
dispersal and logistic growth with constant rate in partially filled demes.

We will argue in this article that the BSC
emerges generically in models of rapidly adapting asexual populations in a
similar way as it describes genealogies in traveling waves of FKPP type.
We present extensive computer simulations and investigate the distribution of
heterozygosity in the population, the average time to the most recent common
ancestor, and the site frequency spectrum (SFS). Most notably, the SFS is 
non-monotonic with a large number of high frequency
derived alleles. We then study a simplified model analytically and 
show that the underlying genealogical process is approximately the
BSC. In the discussion, we outline the basic features of 
the BSC and discuss its applicability to wider classes of models.

\section{The model}
The evolutionary dynamics of a large population are mainly determined by the distribution
of fitness in the population. In general, fitness depends on many traits, which are affected
by mutations. In a rapidly changing environment, populations are far from
any fitness optimum, with many mutations available that increase fitness (and even more
that decrease fitness).

\begin{figure}[thb]
\begin{center}
  \includegraphics[width=0.9\columnwidth]{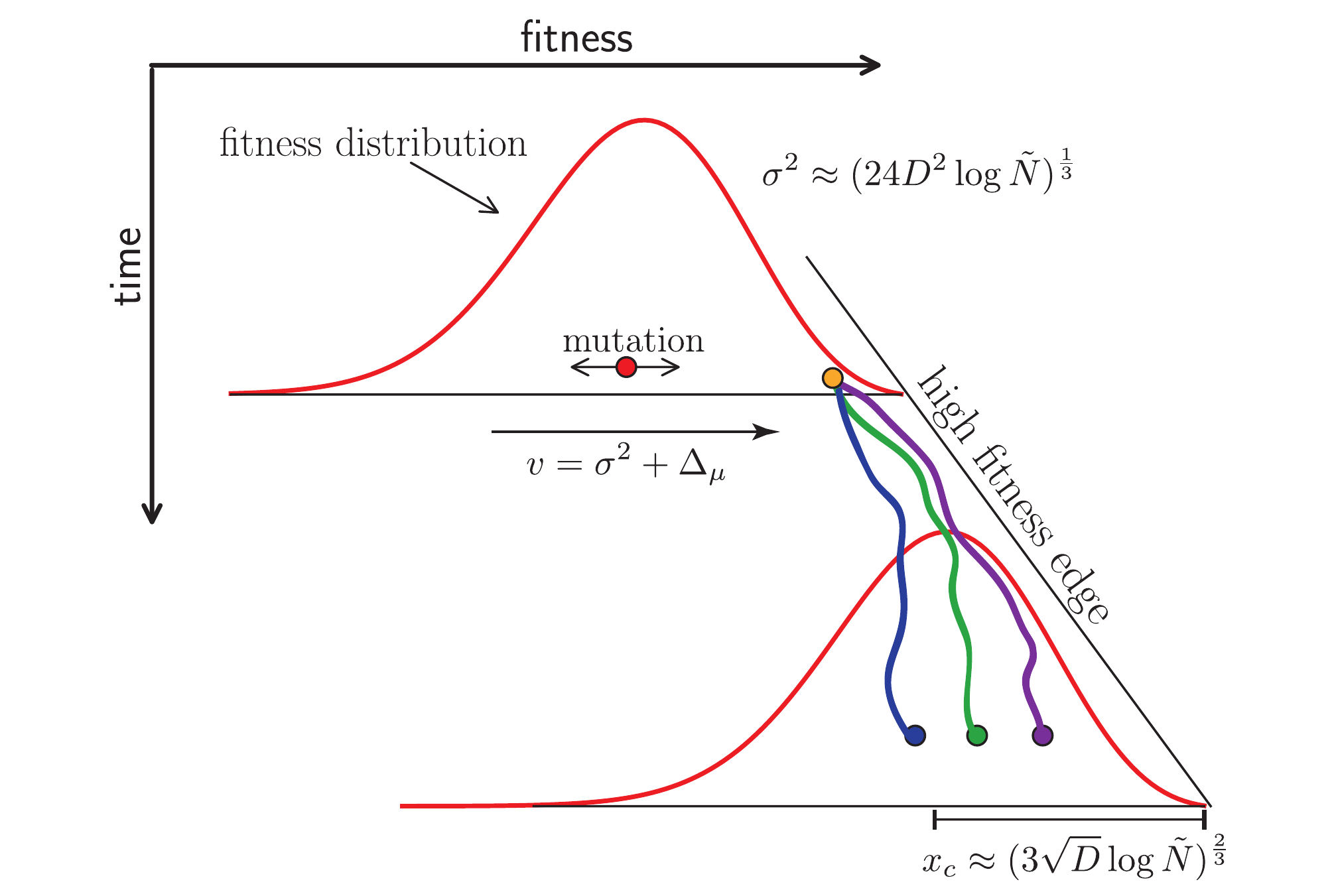}
  \caption{
  Ancestral lineages in evolving populations. The figure shows 
  the fitness distribution of the population, translating towards higher fitness
  with velocity $\sigma^2+\avgmut$, at two time points.
  Randomly sampled individuals (green, blue, and violet dots in the later population)
  tend to come from the center of the distribution, while ancestors tend to be among
  the fittest in the population. The ancestral lineages wiggle due to
  mutations that randomly perturb their fitness. Simultaneously, lineages move towards 
  the high fitness edge, where they are likely to meet and coalesce. The fittest individuals are typically at
  $\x_c\approx \sigma^4/4D$ above the mean fitness.  \label{fig:sketch}}
\end{center}
\end{figure}

To model such scenarios, we consider a collection of  $N$ asexual individuals
that are characterized by a log-fitness $\xu$, which determines their average
reproductive success. Specifically, the number offspring of an individual is
Poisson distributed with mean $\exp(\xu-\lambda)$, where $\lambda = \mfit
-1+N/N_0$ keeps the population size roughly at $N_0$. The log-fitness of
individuals is changed by mutation with probability $\mu$ per generation, where
the mutational effect, $\delta$, is drawn from a distribution $K(\delta)$.
The  balance between frequent mutation and selection results in a  population
that behaves as a traveling pulse along the fitness axis with a steady fitness
variance $\sigma^2$; see \FIG{sketch}.
Absolute fitness itself is  of course not increasing indefinitely, but
increasing fitness is offset by environmental deterioration and deleterious
mutations.

We have implemented this model as a computer program (see SI) that allows for
different mutation distributions $K(\delta)$. In addition, the program keeps
track of the parents of each new individual and thereby saves the complete
genealogy of the population. Individuals not leaving any offspring are removed
from the genealogical record. From this genealogical record, quantities like
pair coalescence times are readily obtained. Furthermore, we can calculate site
frequency spectra of neutral mutations by integrating over all positions in the
genealogies where such mutations might have occurred.

Similar models have been used by a number of authors 
\cite{Tsimring:1996p19688,Desai:2007p954,Rouzine:2003p33590,Neher:2010p30641,Hallatschek:2011p39697,Park:2007p4335}
who have studied the rate of adaptation in these models. Here, we focus on
genealogies and their relation to observed genetic diversity.
If mutations are frequent relative to the typical effect size of mutations, the
model has a continuous time limit described by a stochastic differential
equation for the distribution $c(\xu,t)$ of log-fitness $\xu$ in the population
\cite{Cohen:2005p45154,Good:2012p47545,Hallatschek:2011p39697}
\begin{equation}
\label{eq:model}
\partial_t c(\xu,t) = D\partial_\xu^2 c(\xu,t) - \avgmut \partial_\xu c(\xu,t)  + (\xu-\lambda)c(\xu,t) +
\mathrm{drift} \ ,
\end{equation}
where the last term represents the stochastic nature of reproduction; see SI for
derivation. The diffusion constant and the average mutation input are given by
$D=\mu \langle \delta^2\rangle/2$ and $\avgmut = \mu \langle \delta \rangle$,
respectively, where the average $\langle \ldots \rangle$ is over the
distribution of mutational effects $K(\delta)$. The exact form of the distribution
of mutational effects and the relative importance of deleterious and beneficial
mutations are irrelevant as long as this diffusive approximation is valid (see
SI). Unless otherwise stated, we  use $\mu=1$ and draw mutational effects from a Gaussian
distribution with variance $s^2$ and zero mean.

In this model, large populations attain a steady fitness distribution of roughly
Gaussian shape with variance $\sigma^2 \approx  D^{2\over
3}(24\log\Ne)^{1\over 3}$, where $\Ne = ND^{1\over3}$
\cite{Tsimring:1996p19688,Cohen:2005p45154}. The distribution translates towards 
higher fitness with a velocity $v = \sigma^2 + \avgmut$.
The distribution and its landmarks are sketched in \FIG{sketch}.
It is convenient to measure log-fitness relative to the population mean, $\x
= \xu-\mfit$. The fittest
individuals of the population reside roughly $\x_c \approx \sigma^4/4D$ above the
population mean. Computer programs and analysis scripts are available on the author's website.

\section{Results}
We first present simulation results of our model and contrast the patterns of
genetic diversity of continuously adapting populations with neutral
expectations. Below, we will analyze our model mathematically and show that the
striking differences result from the exponential amplification of individual
lineages by selection. 


\subsection{Distribution of heterozygosity and pair coalescence times}
\begin{figure}[thb]
\begin{center}
  \includegraphics[width=0.95\columnwidth]{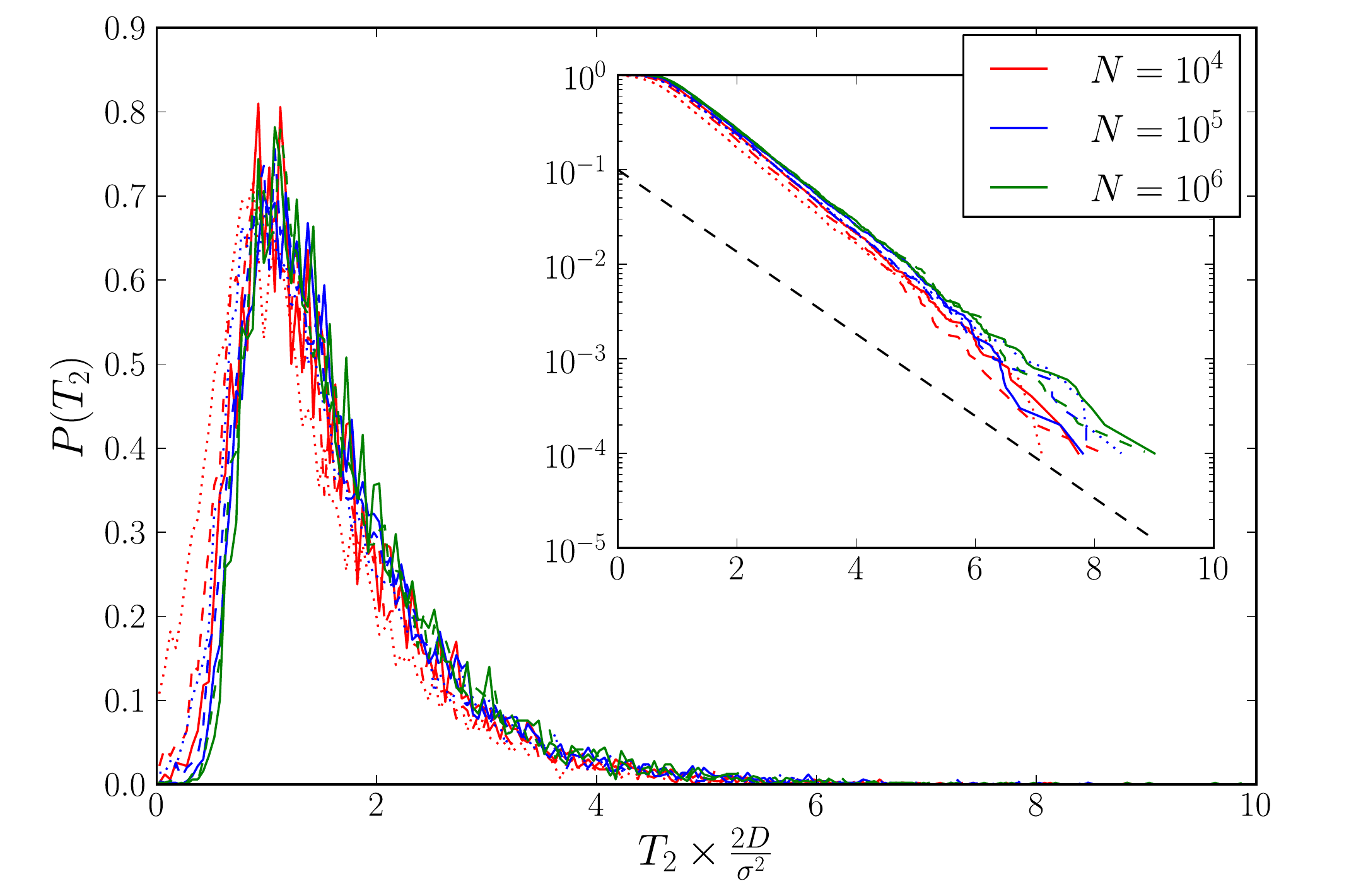}
  \caption{
  The distribution of pair coalescence times (proportional to heterozygosity) in 
  a model of rapidly adapting populations. After rescaling time by $\sigma^2/2D$ curves for different $N$ and $s$  collapse onto a single master curve. This collapse demonstrates
  that $\sigma^2/2D\propto D^{-{1\over 3}}(\log \Ne)^{1\over 3}$ is the time scale of coalescence. Following a delay $T_{delay}\approx \sigma^2/2D$, $T_2$ is exponentially distributed, as is apparent from the inset showing the cumulative 
  distribution $P(T_2>T)$. An exponential $\exp[-2TD/\sigma^2]$ is indicated as a black dashed line. Different line styles correspond to $s=0.01$ (solid),
  $s=0.001$ (dashed), and $s=0.0001$ (dotted), while the mutation rate is $\mu=1$. For each parameter combination, random pairs 
  are sampled at 10000 time points $2s^{-2/3}$ generations apart.
  \label{fig:heterozygosity} }
\end{center}
\end{figure}

Assuming a molecular clock, the expected number of neutral differences between
two genomes is $\pi = 2T_2\mu_n$, where $\mu_n$ is the neutral mutation rate and
$T_2$ is the time to the most recent ancestor of the pair of sequences. Across 
many realizations of the process (e.g.~independent loci), $T_2$
follows a distribution $P(T_2)$, which in the neutral case is exponential with
mean $N$. Simulation results for our model shown in \FIG{heterozygosity} display
a very different distribution of $T_2$ and equivalently $\pi$. Very few
pairs of sequences coalesce early, which results in the long terminal branches observed in trees; see \FIG{influenza}B. 
We then observe a peak in coalescence around $t \approx \sigma^2/2D$,
after which the distribution of pair coalescence times decays exponentially with
a characteristic time constant proportional to $\sigma^2/2D$. Within a neutral
coalescent framework, a distribution of this kind would be interpreted as a
rapid population expansion starting $\sigma^2/2D$ in the past. Prior to this expansion,
the population size would be estimated to have been constant at $N_e \propto
\sigma^2/2D$. However, the size of the population did not change in our model. Instead,
the population was adapting by many small steps, and the conclusion that $N$
increased in the past is wrong.

Two lineages chosen at random from the population are most likely from near the
center of the fitness distribution. There are many individuals in this part of
the distribution, so the probability of immediate coalescence is therefore low.
While the sampled individuals are typical, their ancestors tend to have higher
than average fitness. Only after ancestral lineages have moved to the high
fitness tail of the distribution, where only few individuals are, does the rate
of coalescence become appreciable. This migration of lineages towards higher
fitness is a well known effect
\cite{Hermisson:2002p47231,Ofallon:2010p43263,Rouzine:2007p17401} and
illustrated in \FIG{sketch}.
The speed at which lineages move toward higher (relative) fitness is initially 
$\sigma^2$ (the speed of the mean minus the mutational input), while they slow
down as they reach the tip.
Consistent with the above interpretation, the delay of coalescence, $T_{delay}$,
is roughly twice the time required for the mean fitness to catch up with the
high fitness nose, i.e., $T_{delay} \approx 2x_c \sigma^{-2} = \sigma^2/2D$.
After lineages have moved to the high fitness tail, they seem to coalesce
uniformly with a time constant $T_c \approx \sigma^2/2D$.
From the dependence of $\sigma^2$ on population parameters, we see that
$T_c\propto (\log \Ne)^{1\over 3}$ increases only weakly with the population
size.

\subsection{Site frequency spectra}
\begin{figure}[btp]
\begin{center}
  \includegraphics[width=0.95\columnwidth]{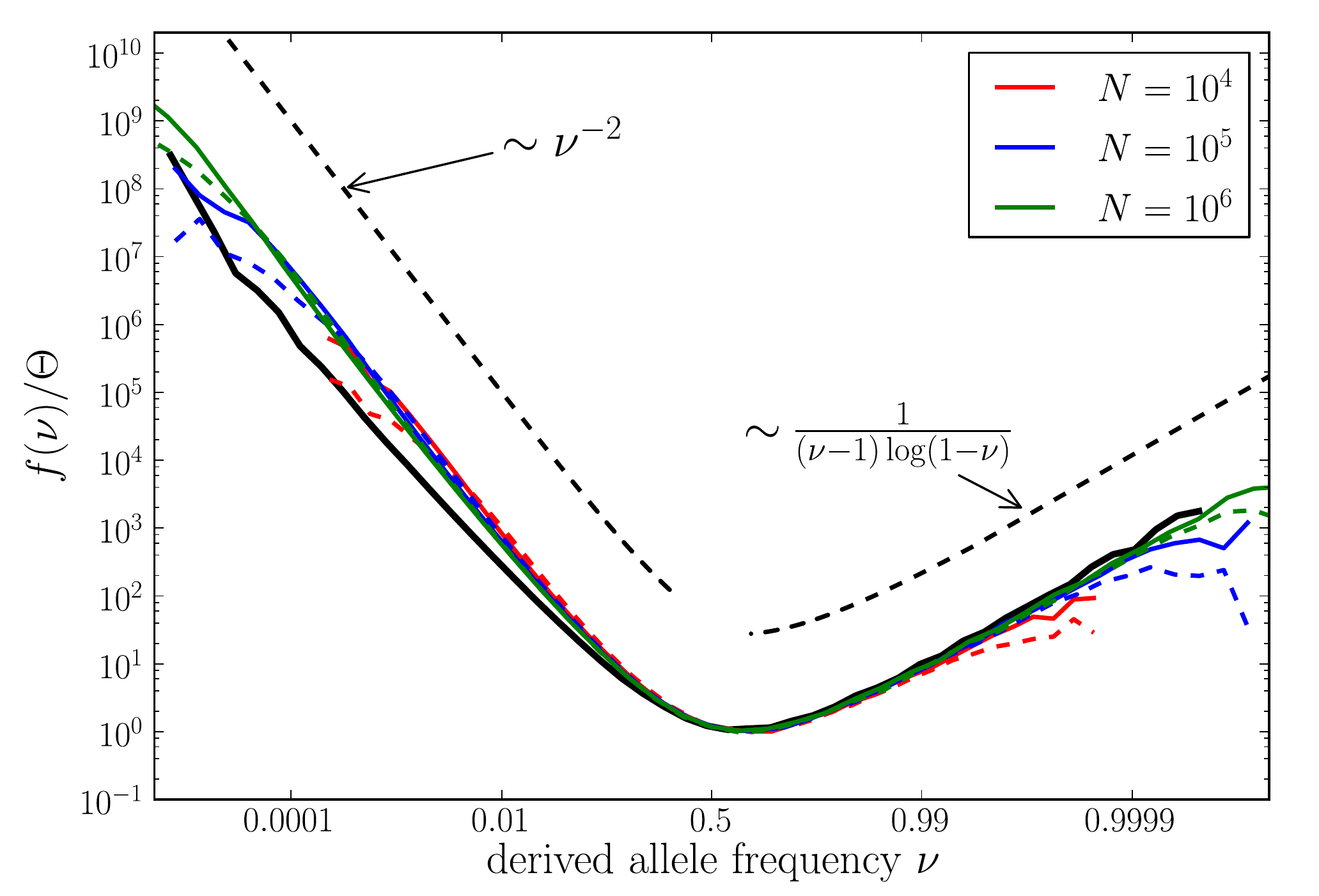}
  \caption{Site frequency spectrum of (derived) neutral alleles in rapidly adapting
  populations is non-monotonic with peaks at low frequencies and near fixation. The asymptotic behavior of the SFS at low and high derived frequencies is shown as dashed black lines.  The solid black line is the SFS of the BSC simulated using \EQS{genBS}{eventrate} with $N=100000$ and averaged over 10000 runs. Line styles and parameters are as in \FIG{heterozygosity}. \label{fig:af}}
\end{center}
\end{figure}

The density $f(\nu)d\nu$ of neutral derived alleles in the frequency interval $[\nu,
\nu+d\nu]$ is known as the site frequency spectrum (SFS). The neutral SFS is a convenient
summary of the neutral diversity segregating in the population. A mutation that
happened on a particular branch of the genealogy will later be present in all
individuals that descend from this branch. Hence the SFS harbors information about
the distribution of branch weights and the branch length of the genealogy.
In Kingman's coalescent, the SFS is simply given by
$f(\nu) = \Theta/\nu$, where $\Theta = 2\langle T_2\rangle\mu$ is the average heterozygosity.
Importantly, it is a monotonically decreasing function of the frequency.
\FIG{af} shows site frequency spectra measured in simulations of our model. The
most striking qualitative difference is the non-monotonicity, a feature known to
be common in the presence of selective sweeps  due to hitch-hiking \cite{Fay:2000p35077}.

The non-monotonicity of $f(\nu)$ implies the existence of long branches
deep in the tree that are ancestors of almost everybody in the population, whereas
a small minority of the population descends from different lineages. Such very
asymmetric branchings are unlikely in Kingman's coalescent, where at any
split the fraction of individuals that go left or right is uniformly
distributed~\cite{Derrida:1991p1707}.
Such asymmetric branchings are common in our model and frequently observed in
reconstructed genealogies from rapidly adapting organisms; see \FIG{influenza}A.

The axes in \FIG{af} are scaled to facilitate the comparison to analytic results.
At low frequencies, the site frequency spectrum is proportional to $\nu^{-2}$
\cite{Basdevant:2008p45781,Neher:2011p42539} and therefore much steeper than
the neutral SFS in Kingman's coalescent, $f(\nu)\propto\nu^{-1}$. Hence samples will be dominated by
singletons. In addition, $f(\nu)$, is non-monotonic and increases as $\nu\to 1$. 

The majority of the contributions to the increase of $f(\nu)$ for $\nu\to 1$
stem from the very last coalescent event. In this last coalescent event, two or
more lineages are merging. One of these lineages is typically the ancestor of
almost the entire sample, while the others share the remaining minority.
The distribution of the offspring of these lineages and their number has been
studied by Goldschmidt and Martin \cite{Goldschmidt:2005p47276}, who showed
that the distribution of the size of the biggest lineage is asymptotically
$\propto (1-\nu)^{-1}$. In the SI (section 3), we derive the more accurate
approximation
\begin{equation}
f(\nu) \approx \frac{T_c\mu}{(\nu-1)\log (1-\nu)} \qquad 1-\nu\ll 1 \ .
\end{equation}
To compare the SFS of our model to that of the BSC across the entire
range of $\nu$, we simulated the idealized BSC and find very good agreement 
(solid black line in \FIG{af}).
The SFS of the idealized BSC deviates from that of the model of adaptation only
at very low allele frequencies. The model of adaptation tends to have even more
rare alleles than the BSC, which is due to the fact that lineages have to move 
to the high fitness tail before coalescence sets in.

The non-monotonicity $f(\nu)$ is a clear indication that the genealogies in this model
with selection are fundamentally different from canonical neutral genealogies (Kingman). 
In Kingman's coalescent, neither constant or exponentially growing population sizes give rise 
to non-monotonic SFS; see supplementary Fig.~\suppfigSFS.

\subsection{The time to the MRCA}
In Kingman's coalescent, the expected time to the most recent common ancestor
(MRCA) of a sample of size $n$, $\langle T_{MRCA}\rangle  = N(2-2/n)$,  increases only very slowly  with $n$.
This is a consequence of the even branching ratios; an additional individual
will most likely coalesce with existing samples and only rarely increases
$T_{MRCA}$.
In contrast, the trees generated by our model of adaptation tend to branch very
unevenly and one often observes that one external branch goes all the way back
to the MRCA of the sample, as in \FIG{influenza}B. As the sample size is increased, 
one continues to sample deeper into the tree. This is a generic property of
the BSC \cite{Berestycki:2009p45808}, where the average $\langle
T_{MRCA}\rangle$  increases as $\log\log n$ with the sample size $n$.
Similarily, $\langle T_{MRCA}\rangle$ of the entire population is expected to
increase as $T_c \log\log \Ne$ with the population size. Our simulations are 
consistent with this behavior; see Fig.~\suppfigTMRCA.

Note that $T_c$ depends weakly on $N$ in adapting populations, while it increases linearly with
$N$ in Kingman's coalescent. In contrast, the rescaled time to the MRCA,
$T_c^{-1}\langle T_{MRCA}\rangle$, asymptotes to $2$ in Kingman's coalescent,
while it continues to increase with $N$ in the BSC.

\section{Analysis}
The simulation results presented above show that genealogies arising in our
model are distinct from those expected in Kingman's coalescent and display
a number of features reminiscent of the Bolthausen-Sznitman coalescent. We will
now describe how this coalescent process emerges from the dynamics of the model.

\begin{figure}[htp]
\begin{center}
  \includegraphics[width=\columnwidth]{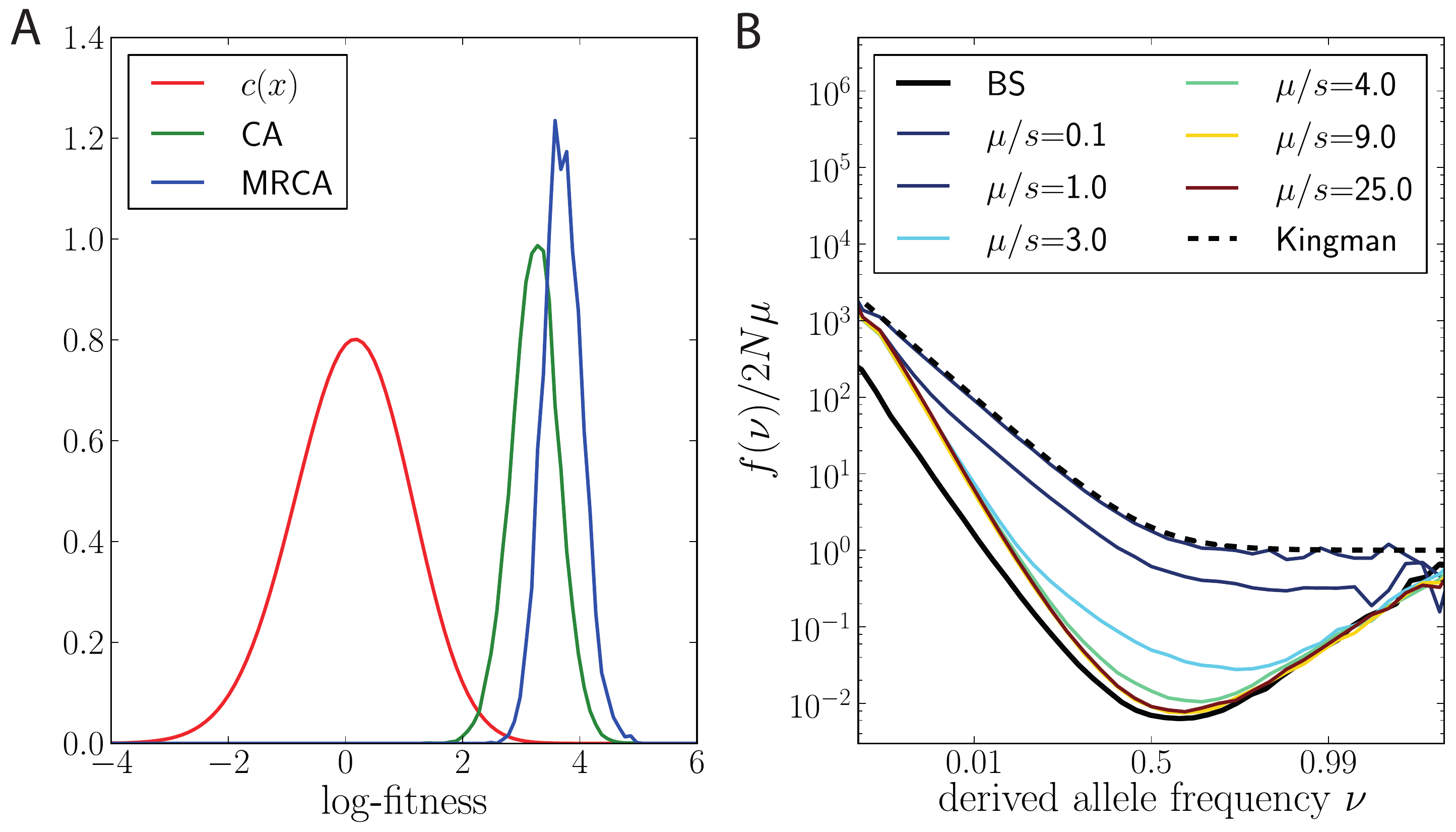}
  \caption{Panel A shows the distribution of the log-fitness of all CAs and all MRCAs compared to the average distribution, $c(\xs)$, of log-fitness in the population; see text. These distributions were measured in forward simulations with $\mu=1,\,s=0.01$ and $N=10^6$. Panel B: Site frequency spectrum of derived neutral alleles in a background selection scenario with deleterious mutations of effect $s$. 
  As the ratio $\mu /s$ is varied while keeping $\sigma^2=\mu s = 0.1$ constant, the SFS interpolates between 
  the expectation for the Kingman and the BS coalescent. $N =10^4$.\label{fig:ratchet} \label{fig:MRCA}}
\end{center}
\end{figure}

Individuals in our model have a heritable fitness which determines the
distribution of the number of immediate offspring. While fit individuals have on
average more offspring than less fit ones, the fitness differences in the
population are small and the offspring distribution across the population is
narrow. However, fitness is heritable and fit individuals can have a very large
number of distant great${}^t$-grandchildren. Hence the distribution of offspring
after $t$ generations, $P(n,t)$, will be dominated by fit individuals and can
have a very long tail. Conversely, the present day population has fewer and
fewer ancestors as we trace its lineages backwards in time.
At $T_{MRCA}$ generations in the past, there is exactly one individual that is the ancestor to the
entire population. Ancestors of the MRCA are also common ancestors (CA) of the
entire population, albeit not the most recent one.
\FIG{MRCA} shows that MRCAs and CAs tend to come from the high fitness tail of
the population.
MRCAs tend to be fitter than CAs, since they are conditioned on giving rise to
at least two lineages that persist to the present.

The offspring distribution, $P(n,t)$, changes slowly from the initial narrow distribution
to a broad distribution with a power-law at intermediate times \cite{Neher:2011p42539}.
The  broad distribution at intermediate times is at the heart of the
correspondence of genealogies in models of adapting populations and the
Bolthausen-Sznitman coalescent.

The BSC assumes that all individuals are exchangeable
and that in every coalescent event a randomly chosen set of lineages merges
into one. Each possible merger event has a specific rate associated with it, and
the rate at which $k$ individuals merge into one common ancestor is  $q_k =
T_2^{-1}(k-1)^{-1}$
\cite{Berestycki:2009p45808} (the general expression for the rates is given below in \EQ{genBS}). In
contrast, in the neutral coalescent, higher order coalescence is very rare
$\propto N^{-(k-1)}$. We will present the basic properties of the BSC briefly in
the discussion.

To appreciate how these coalescence rates can emerge from a model with
selection, consider the number of individuals $n_i$ that descend from an
individual $i$ that lived $t$ generations in the past. The probability that $k$ individuals
sampled randomly from the population have a common ancestor $t$ in the past is
then given by
\begin{equation}
\label{eq:Qk}
Q_k(t) =\left \langle \sum_{i=1}^N \left(\frac{n_i}{\sum_j n_j}\right)^k \right\rangle \ ,
\end{equation}
where the average $\langle .\rangle$ is over all $n_i$. $Q_k(t)$ is dominated by $n_i\gg k$ such 
that sampling with replacement in \EQ{Qk} 
is an accurate approximation. Using the identity
$\Gamma(k) C^{-k} = \int_0^z dz\, z^{k-1} e^{-zC}$ and assuming that $t$ is small enough that the 
different $n_i$ are still approximately independent, we can express 
$Q_k$ as
\begin{equation}
\begin{split}
Q_k(t) &\approx \frac{N}{\Gamma(k)}\int_0^\infty dz z^{k-1} \langle e^{-zn} \rangle^{N-1}\langle n^k e^{-zn}\rangle \\
&\approx -\frac{N}{\Gamma(k)}\int_0^\infty dz z^{k-1} e^{-N\Phi_z(t)} (-1)^{k}\partial_z^k \Phi_z(t) \ ,
\end{split}
\end{equation}
where we introduced the Laplace transform $1-\Phi_z(t) = \sum_n
e^{-zn}P(n,t)$ and assumed $N\Phi_z^2(t)\ll1$.  
In the SI, we show that $\Phi(t)\sim z^{\frac{\sigma^2}{2Dt}}$ for $t >
T_{delay} = \sigma^2/2D$.
For a limited interval after $t>T_{delay}$, we find that the probability that
$k$ individuals have a common ancestor increases with rate
\begin{equation}
\label{eq:BSrates}
q_k = \frac{2D}{\sigma^2}\frac{1}{k-1}
\end{equation}
per unit time. More general coalescence rates can be calculated analogously; see SI. 
Prior to $T_{delay}$, the rate of coalescence is very low. This is in agreement 
with \FIG{heterozygosity}, where we found that little coalescence happened early, while coalescence
times are exponentially distributed after that with characteristic time $T_c \approx \sigma^2/2D$ for $t>T_{delay}$.
The relative rates of mergers of 2,3,\ldots are consistent with the BSC, explaining our observations for the frequency spectrum and the time to the most
recent common ancestor. 

The branching process approximation used to derive the result \EQ{BSrates} is
valid only for short times but nevertheless gives us the relative rates of
multiple mergers once coalescence sets in. For subsequent deeper coalescent
events, the relevant lineages are already at the tip of the fitness distribution,
and this process repeats itself without the delay.
In fact, after this delay all remaining lineages are in a narrow region at the
tip of the fitness distribution. The situation now
resembles that of coalescence in FKPP waves: The fitness of the lineages is
roughly equal, but lineages have to stay ahead of a fitness cut-off in order to survive. We can therefore
employ the phenomenological theory of genealogies in FKPP waves from
\cite{Brunet:2006p47336}, which confirms the above result for the coalescent
time scale; see SI. We present an additional argument based on ``tuned'' 
models introduced in \cite{Hallatschek:2011p39697} in the SI.

To corroborate our analysis, we performed additional simulations that allow us
to measure the Laplace transform of the distribution
of pair coalescent times for very large populations. These simulations show
that the pair coalescent time is indeed exponential with characteristic time $\sigma^2/2D$
after a delay of the same length; see Fig.~\suppfigLargeN.
The algorithm used is similar in spirit to the algorithm by Brunet et al.~
\cite{Brunet:2007p18866}; see SI. 

Strictly speaking, the analogy to an exchangeable coalescent model
like the Bolthausen-Sznitman coalescent requires that different
coalescence events one lineage undergoes be independent. For this to
be true, individuals descending from a lineage have to distribute
evenly across the fitness distribution $c(\x)$ between coalescence
events, which requires a time $T_{eq}\approx T_c$.  Hence we should
not expect a clean convergence towards the Bolthausen-Sznitman
coalescent.  Nevertheless, we find it to be a very good model for the
observed genealogies after accounting for the delay. The underlying
reason is that local equilibration in the region where the ancestral
lineages are is fast ($t\approx D^{-1/3}$). This region, however,
undergoes fluctuations on the time scale $T_{eq}$, which modulate the
overall rate of coalescence but do not significantly affect the local
dynamics. For waves of FKPP type that describe the spread of
individuals in space, $T_{eq}\ll T_c$ in large populations
\cite{Brunet:2007p18866}.

\section{Discussion}
We have shown that in a simple model of adapting populations, the observed
genealogies are inconsistent with the standard neutral coalescent.
Instead, genealogical trees are characterized by long terminal branches and
almost simultaneous coalescence of multiple lineages. At branching events deep
in the tree, one commonly observes that almost all individuals of the population
descend from one branch, whereas very few descent from the other branches. Such
skewed branching is unlikely in neutral coalescent models, regardless of the
history of the effective population size. One consequence of these uneven
branching ratios is a non-monotonic site frequency spectrum (SFS) of derived
neutral alleles. Compared to the neutral coalescent, the low frequency part of
the SFS is much steeper, whereas the high frequency part shows a characteristic
up-turn; see \FIG{af}.

A given pair of lineages is unlikely to coalesce in the bulk of the fitness
distribution. Typically, both lineages move into the high fitness tip of the
population distribution before they coalesce as illustrated in \FIG{sketch}. 
This results in long terminal branches and 
a distribution of heterozygosities peaked at intermediate values. 
After this delay, the typical time to coalescence is
again on the order of the time it takes the fittest individuals to dominate the
population; see \FIG{heterozygosity}. In panmictic populations, this time depends 
on the logarithm of the population size and in our model is proportional to $(\log \Ne)^{1\over 3}$. 

We argue that the exponential amplification of fit lineages is responsible for
these observations and that coalescence in such rapidly adapting populations is
generically described by a modified Bolthausen-Sznitman coalescent (BSC)
\cite{Bolthausen:1998p47390,Berestycki:2009p45808}. The BSC is a special case
of the large class of $\Lambda$-coalescent processes \cite{Pitman:1999p41045}.
Given the distribution $p(f)$ of the fraction $f$ of the population that
descends from a single individual in the previous generation, the rate at which
$k$ out of $b$ lineages merge is given by
\begin{equation}
\begin{split}
\lambda_{b,k}& = \int df p(f) f^k (1-f)^{b-k}\ .
\end{split}
\end{equation} 
The Bolthausen-Sznitman coalescent corresponds to $p(f)\sim f^{-2}$ for large $f$, in which case \EQ{genBS}
reduces to 
\begin{equation}
\label{eq:genBS}
\lambda_{b,k} = \frac{1}{T_c} \frac{(k-2)!(b-k)!}{(b-1)!}
\end{equation}
or \EQ{BSrates} for the special case $b=k$. 
The total rate at which coalescence events happen in a sample of $k$ lineages is therefore
\begin{equation}
\label{eq:eventrate}
\lambda_{b} = \sum_{k}{b\choose k} \lambda_{b,k} = \frac{b-1}{T_c} \ ,
\end{equation}
in contrast to the neutral coalescent, where $\lambda_b \propto b(b-1)$. 
A coalescence event reduces the number of surviving lineages on average by $\log b$ such that
the average rate at which the number of lineages decreases is $\approx T_c^{-1} b\log b$. 
The typical time needed to reach the common ancestor of a sample of size $n$ is $\approx T_c\log\log n$, in
contrast to $2T_c$ in Kingman's coalescent. The BSC occupies a special intermediate position
between Kingman's coalescent, where only pairwise mergers are allowed, and a star-coalescent, where
all lineages coalesce simultaneously. Star-like genealogies are expected in rapidly expanding
populations or in a region fully linked to a recent rapid hard sweep \cite{Slatkin:1991p43283}.
In the BSC, multiple mergers (subsets of lineages with star-like trees) are frequent, 
but at the same time there are many mergers at different depth in the tree. In fact, the 
BSC is the $\alpha=1$ case of the one parameter family of Beta-coalescents with parameter $0<\alpha<2$,
while the case $\alpha\to 2$ corresponds to the Kingman coalescent. For a more in depth discussion, 
see the recent review by Berestycki \cite{Berestycki:2009p45808}.
The BSC is easily implemented as a computer simulation by drawing an exponentially distributed
random number with mean $\lambda_b^{-1}$ to determine the time of the next event. The type of event
is then chosen with probabilities proportional to $\lambda_{b,k}$. 

The models of adaptation we have studied have a narrow offspring distribution.
Nevertheless, the exponential amplification of fit genotypes over many
generations gives rise to a distribution of clone sizes with the required
asymptotic behavior. The important lineages are those that run ahead of
the distribution, expand faster, and take over a significant fraction of 
the population~\cite{Brunet:2006p47336}. Over even longer times, the fitnesses of ancestors and 
descendants decorrelate. This allows us to approximate the genealogies with 
the abstract BSC, which assumes that there are no 
correlations in offspring number across generations. 

Conventionally, an increased variance in offspring number is accounted for by
defining an effective population size. With a clone size distribution $p(f)\sim
f^{-2}$, however, the variance diverges with the population size
\cite{Neher:2011p42539}. Similar effects arise in other models with very skewed
offspring distribution \cite{Eldon:2006p36003}. As a consequence, the genealogies are dominated by
rare anomalously large clones and described by the BSC rather than Kingman's
coalescent. The rate of coalescence is not set by $N^{-1}$ but by the
rate at which clones expand and collapse. We would like to
stress that evolutionary dynamics thereby remains highly stochastic, even in
very large populations. Analogous behavior has recently been observed in models
of individuals invading uninhabited territory (FKPP type waves)
\cite{Brunet:2007p18866} and ensembles of super-critical branching processes
\cite{Schweinsberg:2003p40932}.

The BSC is not only a good model for genealogies of adapting asexual populations but
also applies to populations under purifying selection in which Muller's ratchet
clicks often. The standard model for the distortion of genealogies by purifying
selection assumes that deleterious mutations are rapidly purged and coalescence
is neutral in the mutation free class with a reduced population size $Ne^{-\mu/s}$, 
where $\mu$ is the deleterious mutation rate and $s$ is the effect size of deleterious
mutations \cite{Charlesworth:1993p36005}. More elaborate analysis based on a
fitness-class coalescent explicitly tracks lineages through the population and
calculates the contribution to coalescence before lineages reach the mutation
free class \cite{Walczak:2011p45228}. 
However, all of this only holds as long as the mutation free class is large and
Muller's ratchet does not operate, which requires $N s e^{-\mu/s}\gg 1$ \cite{Stephan:1993p42929,Jain:2008p45047,Neher:2012p47353}.
Fig.~\ref{fig:ratchet}B shows the SFS of derived neutral alleles for different ratios $\mu/s$.
For small $\mu/s$, the SFS is similar to those of Kingman's coalescent with a
reduced time to coalescence, in accordance with the background selection theory.
However, as soon as the ratchet starts to click frequently, the SFS develops the
non-monotonicity characteristic of the Bolthausen-Sznitman coalescent.

If the ratchet is clicking fast, the fitness distribution in the population resembles that of
traveling wave models, but selection on fitness variation can not keep up with
the influx of deleterious mutations. Similarly, populations in a steady balance
between deleterious and beneficial mutations \cite{Goyal:2012p47382} have
genealogies as found here for rapidly adapting populations. The reason for the
qualitative difference in the ratchet regime is the fact that the nose of the wave is not steady, but
constantly turning over. Different lineages are struggling to get ahead of
everybody else and, in the frame of reference of the population (that is,
relative to mean fitness), are exponentially amplified. In contrast, dynamics of
lineages in the mutation free class is neutral if Muller's ratchet does not
operate.

In the supplementary material, we show that the argument that gave rise to the
particular coalescence rates in \EQ{BSrates} can be extended to a large class of
models that are controlled by a small and fluctuating population of highly fit
individuals. We thus argue that the Bolthausen-Sznitman coalescent generically
emerges as a consequence of the exponential amplification of the clones
descending from these highly fit individuals, together with the seeding of
novel lineages.  The latter could happen via lucky diffusion to high fitness
(our model here), via large effect beneficial mutations, or via lucky
outcrossing. After some time, the distribution of lineage size follows a
power law with an exponent close to -2
\cite{Yule:1925p38579,Desai:2007p954,Neher:2011p42539}.
Given an effective offspring distribution of this shape, the Bolthausen-Sznitman
coalescent follows \cite{Schweinsberg:2003p40932,Brunet:2007p18866}. In
\cite{desai_genetic_2012}, the authors study a model where the mutation rate 
is much smaller than the typical effect sizes of mutations. They show that also in this
case, the genealogies are well approximated by the BSC after a delay. Whether the BSC also
describes genealogies in scenarios where fitness is increased in rather large
increments (compared to the population diversity)
\cite{Gerrish:1998p5933,Schiffels:2011p43799} remains an interesting topic for
future work.

The compatibility of a sample with the neutral coalescent model is
typically assessed using statistics such as Tajima's $D$
\cite{Tajima:1989p37563}. Tajima's $D$ compares the average number of
pairwise differences to the total number of segregating sites in the
sample. In the case of the BSC, the average
pairwise diversity is proportional to $\langle T_2\rangle$, while the total number of
segregating sites is proportional to $\langle T_2\rangle n/\log n$ (compared to
$\langle T_2\rangle\log n$ for the Kingman coalescent). This tremendous excess of
segregating sites is a consequence of the very steep SFS at small
frequencies and results in $D\propto -\log(n)$. 

Sexual populations and recurrent selective sweeps at linked loci can also give rise to
multiple mergers in the genealogies \cite{Durrett:2005p40919}. However,
recombination and sexual reproduction will reduce the effects of linked
selection and decouple the genealogies of different loci. Hence we expect that
the coalescent behavior crosses over to Kingman's coalescent as the
recombination rate increases -- at least in models of panmictic populations.
This is indeed observed in models of facultatively sexual populations
\cite{Neher:2011p42539}.

Given the apparent universality of the Bolthausen-Sznitman coalescent in spatially 
expanding populations and panmictic adapting populations, it should be included as
a prior in popular population genetic and phylogenetic inference programs 
such as BEAST \cite{Drummond:2007p30127}.

\section{Acknowledgements}
We are grateful for many stimulating discussions with Boris Shraiman, Aleksandra
Walczak, Michael Desai, Daniel Fisher, Trevor Bedford, and Martin M\"ohle. We
also would like to thank Kari K\"{u}ster for coding some of the simulations used
in early stages of this work and Lukas Geyrhofer for help with tuned models. 
This research was supported by the ERC through Stg-260686 to RAN.

\clearpage
\appendix
\onecolumngrid
\section{Derivation and limitations of Eq.~1}
\label{sec:appEq1}
We assume a population of $N$ individuals that replicate at rate $B$, die at
rate $D$, and mutate at rate $\mu$. For convenience, we measure time in units of
generations and set birth and death rate to $B=\exp(\xu-\lambda)$ and $D=1$,
respectively. Here, $\xu$ is the log-fitness of the individual in question,
while the population size is kept constant by adjusting $\lambda$.
Typically, $\lambda$ is equal to the sum of the mean log-fitness and a small
modulation that increases growth rate when the population size falls below the
target $N_0$ and decreases growth rate otherwise. We use $\lambda =
\mfit-(1-NN_0^{-1})$ which keeps $N$ within $\mathcal{O}(\sqrt{N_0})$ of $N_0$.
For present purposes, this model is equivalent to a model with strictly constant
population size. Different forms of $\lambda$ that constrain the population size
to a different degree are possible \cite{Hallatschek:2011p39697}.
Mutations increment the log-fitness $\xu$ of an individual by $\delta$ drawn
from a distribution $K(\delta)$.
Disregarding fluctuations and assuming fitness differences are small, the
distribution of fitness in the population, $c(\xu, t)$, obeys the following
deterministic equation
\begin{equation}
\label{eq:apppopdis}
\partial_t c(\xu,t) = (\xu-\lambda)c(\xu) + \mu \int ds\, K(\delta)\left[c(\xu-\delta,t)-c(\xu,t)\right]
\end{equation}
Generically, steady state solutions to this equation oscillate around
$0$ for large $\xu-\mfit$, where the stochastic effects of
finite populations become important \cite{Cohen:2005p45154}. 
The magnitude of stochastic
perturbations to the population in the interval $\left[\xu,
  \xu+\Delta\xu \right]$ are proportional to $\sqrt{c(\xu)\Delta\xu}$
and can be accounted for either by an explicit model of discrete
particles, or a stochastic partial-differential equation
\begin{equation}
\partial_t c(\xu,t) = (\xu-\lambda)c(\xu) + \mu \int d\delta\, K(\delta)\left[c(\xu-\delta,t)-c(\xu,t)\right] + \sqrt{c(\xu,t)} \eta(\xu,t)
\end{equation}
where $\eta(\xu,t)$ is Gaussian white noise with $\langle \eta(\xu,t) \eta(\xu+x,t') \rangle = \delta(x)\delta(t-t')$
\cite{Hallatschek:2011p39697}. This stochasticity determines the average steady
state velocity, which diverges in the limit $N\to \infty$
\cite{Rouzine:2003p33590,Desai:2007p954,Cohen:2005p45154}.

If the mutation rate $\mu$ is large and the distribution $K(\delta)$ falls off
quickly enough such that the typical mutations are small, one can expand
$c(\xu-\delta,t)$ inside the integral to obtain
\begin{equation}
\label{eq:stochPDE}
\partial_t c(\xu,t) \approx  (\xu-\lambda)c(\xu,t) + \frac{\mu \langle \delta^2\rangle}{2}\frac{\partial^2}{\partial \xu^2}c(\xu,t) - \mu \langle \delta\rangle\frac{\partial}{\partial \xu}c(\xu,t) + \sqrt{c(\xu,t)} \eta(\xu,t)
\end{equation}
where $\langle \ldots\rangle$ denotes the average over $K(\delta)$. The expansion of
$c(\xu-\delta,t)$ is a good approximation  as long as the mutation rate is larger than typical 
mutational effects. Specifically, it is required that $\mu^2/\langle \delta^2\rangle > (\log
\Ne)^2$, see \cite{Cohen:2005p45154} for a discussion of this issue.
This condition will limit the applicability of the model to RNA
virus populations or asexual eukaryotes with long genomes. We find, however,
that the basic features of the BSC are also present in models where
$\mu/\sqrt{\langle \delta^2\rangle} \ll 1$, see supplementary figures \ref{fig:supp_af_step} and \ref{fig:supp_af_exp} below. This observation is confirmed by 
a recent preprint by Desai et al \cite{desai_genetic_2012}, in which the authors investigate a model with $\mu^2\ll \langle \delta^2\rangle$.

The average of the population distribution governed by \EQ{stochPDE} obtains a
steady shape that translates towards higher $\xu$ with a velocity $v = \sigma^2
+ \mu\langle \delta\rangle$, where
\begin{equation}
\sigma^2 \approx  D^{2\over 3} \left(24 \log (ND^{1\over 3})\right)^{1\over 3}
\end{equation}
and $D = \frac{\mu \langle \delta^2\rangle}{2}$. In a suitably chosen comoving
frame $\x = \xu-vt$, the equation describing the population distribution
simplifies to
\begin{equation}
\label{eq:appstochPDE}
0 \approx  \x c(\x,t) + D\frac{\partial^2}{\partial \x^2}c(\xu,t) + \sigma^2\frac{\partial}{\partial \x}c(\x,t) + \sqrt{c(\x,t)} \eta(\x,t)
\end{equation}
If the mutational input $\mu\langle \delta\rangle=0$, the average velocity equals
the fitness variance $\sigma^2$ of the population. If deleterious mutations are
more common than beneficial mutations and on average reduce fitness, the
velocity will be smaller than the fitness variance by $|\mu\langle \delta
\rangle|$.
\EQ{appstochPDE} has been studied in detail in \cite{Cohen:2005p45154}. In
large population, the typical solution in the quasi-steady state is given by
\begin{equation}
c(x) = \begin{cases} 
C e^{-\frac{\sigma^2 \x}{2D}} \Ai\left(\frac{\sigma^4}{4D^{4\over 3}}-\frac{x}{D^{1\over 3}}\right) & \x<\x_c \\
De^{-\frac{\sigma^2 \x}{D}} & \x>\x_c
\end{cases}
\end{equation}
where $\x_c$ and $D$ are determined by matching $c(\x)$ and $\partial_\x c(\x)$ at $\x_c$, while $C$
is set by the normalization.

\section{Frequency spectrum of derived alleles in the BSC}
A mutation occurring on some branch of the genealogy will be present in all
individuals that descend from the branch considered. Very recent mutations tend to be rare, 
while old mutations can be common. The number of leafs descending from lineages at a time 
$\tau = T/T_c$ in the past is also known as the size of the $\tau$-families ($T_c$ is the coalescent time scale).
These family sizes follow a two parameter Poisson-Dirichlet distribution $PD(0, e^{-\tau})$
and the size $z$ of the first block is Beta distributed
as~\cite{Berestycki:2009p45808}
\begin{equation}
P(z, \tau) =
\frac{z^{-e^{-\tau}}(1-z)^{e^{-\tau}-1}}{\Gamma(1-e^{-\tau})\Gamma(e^{-\tau})}
\end{equation}
Common mutations, i.e., those present in the majority of the population, most
likely fall onto this block. The present day frequency of the derived allele is equal to the block size. 
If neutral mutations occur at rate $\mu$, we can calculate the ensemble averaged site frequency 
spectrum (SFS) by integrating over all times at which the mutation could have occurred. Ignoring
mutations that fall onto other blocks of the Poisson-Dirichlet distribution, the average 
SFS is given by
\begin{equation}
\begin{split}
f(\nu) &= \mu\int_0^\infty d T
\frac{\nu^{-e^{-\tau}}(1-\nu)^{e^{-\tau}-1}}{\Gamma(1-e^{-\tau})\Gamma(e^{-\tau})}
\\
&= \mu T_c \int_0^\infty d \tau
\frac{\nu^{-e^{-\tau}}(1-\nu)^{e^{-\tau}-1}}{\Gamma(1-e^{-\tau})\Gamma(e^{-\tau})}
\\&= \mu T_c\int_0^1 dy
\frac{\nu^{-y}(1-\nu)^{y-1}}{y\Gamma(1-y)\Gamma(y)}
\\
&= \mu T_c\int_0^1 dy
\frac{\sin(\pi y) \nu^{-y}(1-\nu)^{y-1}}{\pi y}
\end{split}
\end{equation}
where we have used  Euler's reflection formula for the Gamma functions. 
The last integral has an expression in terms of special functions. However, the 
term $\sin(\pi y)/\pi y$ is close to one for small $y$ where the dominant
contribution to the integral comes from. Hence we have approximately
\begin{equation}
\label{eq:appSFS}
f(\nu) \approx \mu T_c \int_0^1 dy \nu^{-y}(1-\nu)^{y-1} = \mu T_c\frac{2\nu-1}{\nu(1-\nu)(\log \nu - \log
(1-\nu))}
\end{equation}
Higher order corrections to \EQ{appSFS} can be obtained by expanding the sine to higher order. 
They do not change the result qualitatively. For common derived alleles with $1-\nu\ll 1$, the above 
restriction to the first block of the Poisson-Dirichlet distribution is a good approximation since other blocks are almost always short $z \ll 1$.  In the limit $1-\nu\ll 1$, the expression simplifies to 
\begin{equation}
\label{eq:appSFSapprox}
f(\nu) \approx  -\frac{\mu T_c}{(1-\nu) \log (1-\nu)}
\end{equation}
which coincides with Eq.~(2) of the main text. For $1-\nu\ll 1$, it
describes the simulation data very well, see \FIG{af}. At the
rare end of the spectrum, the observed power law $f(\nu)\sim \nu^{-2}$ is
a (well-known) consequence of exponentially growing sub-populations \cite{Yule:1925p38579},
corresponding to expanding clones in the high-fitness tip of the traveling
wave.

\section{Generating function of the offspring number distribution}
To calculate the moments of the lineage size distribution, we need the
distribution $P(n,t|\x,t_0)$ of lineage sizes $n$ at time $t$, given the
lineage was seeded a fitness $\x$ above the mean at time $t_0$. $P(n,t|\x,t_0)$ obeys the backward equation 
\begin{equation}
\begin{split}
P(n,t|\x+v\Delta t,t_0-\Delta t) =& \left[1-\Delta t (2+\x+\mu) \right]P(n,t|\x,t_0) + \Delta t (1+\x)\sum_{n'}P(n',t|\x,t_0)P(n-n',t|\x,t_0) \\
&+\Delta t \delta_{n,0}+ \Delta t\mu \int ds\, K(\delta) P(n,t|\x+\delta,t_0)
\end{split}
\end{equation}
where $K(\delta)$ is a normalized distribution of mutational effects as used in Sec.~\ref{sec:appEq1} and we have used $1+x$ as birth rate and $1$ as death rate. The first term gives the probability that no birth, death, or mutation is happening in the interval $\Delta t$, the second term corresponds to birth, the third term to loss of the lineage ($n=0$) and the last term to a mutation changing the log-fitness by $\delta$. Rearranging and taking the limit $\Delta t \to 0$ yields
\begin{equation}
\begin{split}
-\partial_{t_0} P(n,t|\x,t_0)  + v\partial_x P(n,t|\x,t_0) =& -(2+\x+\mu)P(n,t|\x,t_0) + (1+\x)\sum_{n'}P(n',t|\x,t_0)P(n-n',t|\x,,t_0) \\
&+\delta_{n,0}+ \mu \int d\delta\, K(s) P(n,t|\x+\delta,t_0)
\end{split}
\end{equation}

Using the time translation invariance, $P(n,t|\x,t_0)=P(n,t-t_0|\x,0)$, and defining $\psi_z(\x,t) = \sum_n e^{-zn} P(n,t|\x, 0 )$, we find
\begin{equation}
\partial_t \psi_z(\x,t) + v\partial_x \psi_z(\x,t) = 1-(2+\x)\psi_z(\x,t) + (1+\x)\psi_z^2(\x,t) + \mu \int ds\, K(\delta)[\psi_z(\x+\delta,t)-\psi_z(\x,t)]
\end{equation}
Upon multiplying by $-1$ and substituting $\phi_z(\x,t) = 1-\psi_z(\x,t)$, this equation simplifies to
\begin{equation}
\partial_t \phi_z(\x,t) + v\partial_x \phi_z(\x,t) = \x \phi_z(\x,t) - \phi_z^2(\x,t) + \mu \int ds\, K(\delta)[\phi_z(\x+\delta,t)-\phi_z(\x,t)]
\end{equation}
where we have approximated $(1+\x)  \phi_z^2$ by $ \phi_z^2$, which amounts to assuming that all relevant $\x\ll 1$ and selection is important only over many generations. Note the similarity of this equation to \EQ{apppopdis} governing the fitness distribution of the population. In analogy to Sec.~\ref{sec:appEq1} we can expand $\phi_z(\x+\delta,t)$ inside the integral if the mutation rate is large and $K(\delta)$ falls of rapidly with $|\delta|$ and obtain
\begin{equation}
\partial_t \phi_z(\x,t) = \x \phi_z(\x,t) - \sigma^2 \partial_\x \phi_z(\x,t)  + \frac{\mu \langle \delta^2\rangle}{2} \partial_\x^2 \phi_z(\x,t) - \phi_z^2(\x,t)
\end{equation}
where $\sigma^2 = v - \mu\langle \delta\rangle$ accounts for both the actual translation of the population along the fitness axis and the mutation load. Defining $D= \frac{\mu \langle \delta^2\rangle}{2}$, we have the following equation for the generating function
\begin{equation}
\label{eq:appBP}
\partial_t \phi_z(\x,t) = \x\phi_z(\x,t)  - \sigma^2\partial_\x \phi_z(\x,t) +D\partial_{\x}^2 \phi_z(x,t) - \phi^2_z(\x,t)
\end{equation}
with initial condition  $\phi_z(\x,0) = 1-e^{-z}$. This equation allows two
approximate solutions for small and large $\x$.  For small $\x$, $\phi_z(\x,t)$
is small we can neglect the non-linearity. Conversely, at large $\x$, we find
that $\phi_z(\x,t)$ is large and the dominant balance is between the terms $\x
\phi_z(\x,t)$ and $\phi_z^2(\x,t)$ while the first and second derivative can be
neglected.
Using these approximations and assuming $z\ll 1$, we can solve \EQ{appBP} for small $\x$ and large
$\x$ and match the two asymptotic solutions.
\begin{equation}
\phi_z(\x,t) \approx  \begin{cases}
ze^{\x t + \frac{Dt^3}{3}-\frac{\sigma^2 t^2}{2}} & \x<\x^* \\
\x & \x>\x^*
\end{cases}
\end{equation}
The crossover is approximately at $\x^* = \frac{\sigma^2 t}{2}-\frac{Dt^2}{3}
-\frac{\log z/x^*}{t}$. Note that the branching process description is only valid
for a time before population size constraints have to be
imposed and lineages cease to be approximately independent \cite{Cohen:2005p45154,Hallatschek:2011p39697}. 
At short times, however, it is a valid approximate description of the number of offspring of a marked individual.
We are interested foremost in the
distribution of offspring averaged over the population $c(\x)$, see \EQ{appstochPDE}.
For short times with $\x_c<\x^*$ we have
\begin{equation}
\Phi(z,t) = zAe^{-\frac{\sigma^2t^2}{2}+\frac{Dt^3}{3}}\int^{\x_c} d\x \Ai\left(\frac{\sigma^4}{4D^{4\over 3}}-\frac{x}{D^{1\over 3}}\right) e^{x\left(t-\frac{\sigma^2}{2D}\right)}
\end{equation}
We see that the behavior of this average changes qualitatively as
$t>\frac{\sigma^2}{2D}$ when it starts to be boundary dominated. Assuming $\x^*<x_c$,
i.e., the saturation of $\phi_z(\x,t)$ happens within the validity of the Airy
function solution, we have
\begin{equation}
\Phi(z,t) \approx  zAe^{-\frac{\sigma^2t^2}{2}+\frac{Dt^3}{3}}\int^{\x^*} d\x \Ai\left(\frac{\sigma^4}{4D^{4\over 3}}-\frac{x}{D^{1\over 3}}\right) e^{x\left(t-\frac{\sigma^2}{2D}\right)}+x^* A\int_{\x^*}^{x_c} d\x \Ai\left(\frac{\sigma^4}{4D^{4\over 3}}-\frac{x}{D^{1\over 3}}\right) e^{-\frac{x\sigma^2}{2D}}
\end{equation}
Both of these integrals are dominated by a narrow region in the vicinity of $\x^*$. Plugging in the definition of $\x^*$, we find
\begin{equation}
\begin{split}
\Phi(z,t) &\approx  x^* A e^{-\frac{x^*\sigma^2}{2D}}\left[\int^{0} d\delta\x \Ai\left(\frac{\sigma^4}{4D^{4\over 3}}-\frac{x^*+\delta x}{D^{1\over 3}}\right) e^{\delta x (t -\frac{\sigma^2}{2D})}+\int_{0} d\delta \x \Ai\left(\frac{\sigma^4}{4D^{4\over 3}}-\frac{x^*+\delta x}{D^{1\over 3}}\right) e^{-\frac{\delta x \sigma^2}{2D}}\right]   \\
&= C(t,x^*) e^{-\frac{x^*\sigma^2}{2D}} = C(t,x^*) e^{\frac{\sigma^2\log z/x^*}{2D t} + \frac{\sigma^4t}{4D}-\frac{\sigma^2t^2}{6}} = E(t,x^*)z^{\frac{\sigma^2}{2Dt}}
\end{split}
\end{equation}
where $C(t,\x^*)$ and $E(t,\x^*)$ depend only weakly on $x^*$.
The $k$th derivative of $\Phi(z,t)$ with respect to $z$ is therefore 
\begin{equation}
\partial_z^k\Phi(z,t) \approx   z^{-k} \Phi(z,t)\prod_{i=1}^{k-1}(\frac{\sigma^2}{2Dt}-i) = (-1)^k z^{-k} \Phi(z,t)\prod_{i=0}^{k-1}(i-\frac{\sigma^2}{2Dt}) 
\end{equation}
We can now evaluate the probability $Q_k$ that a set of $k$ lineages merges into one:
\begin{equation}
\begin{split}
Q_k(t)  &\approx- \frac{N}{\Gamma(k)} \int_0^\infty dz z^{k-1} e^{-N\Phi(z,t)}(-1)^k \partial_z^k \Phi(z,t)  \\
&\approx -\frac{\prod_{i=0}^{k-1}(i-\frac{\sigma^2}{2Dt})}{\Gamma(k)} \int_0^\infty dz z^{-1} e^{-N\Phi(z)}N\Phi(z,t)\\
&\approx -\frac{2Dt}{\sigma^2} \frac{\prod_{i=0}^{k-1}(i-\frac{\sigma^2}{2Dt})}{\Gamma(k)} \int_0^\infty d\phi\, e^{-\phi} = \frac{\prod_{i=1}^{k-1}(i-\frac{\sigma^2}{2Dt})}{\Gamma(k)} \ .
\end{split}
\end{equation}
Defining $\tau$ as $\tau=\frac{2D t }{\sigma^2}$, we have
\begin{equation}
Q_k(t) = \frac{\Gamma(k-\tau^{-1})}{\Gamma(1-\tau^{-1})\Gamma(k)} \ .
\end{equation}
For large $\tau$, all $Q_k$ become equal, which simply means that the population
has coalesced into its common ancestor. For $t=\frac{\sigma^2}{2D}+\delta t$, we can
differentiate with respect to $\delta t$ and obtain
\begin{equation}
q_k=\partial_t Q_k(t)|_{t=\sigma^2/2D} = \frac{2D}{\sigma^2}\frac{1}{k-1} \ ,
\end{equation}
which are the Bolthausen-Sznitman merger rates.

Similarly, we can calculate more general merger probabilities such as the probability that $k$ out of $b$
lineages have merged, while all other $b-k$ lineages trace back to distinct ancestors. Making again the assumption of independent
lineages and accounting for the possibilities of choosing $b-k+1$ ordered lineages as ancestors, we have
\begin{equation}
\begin{split}
\Lambda_{b,k} &= \frac{\Gamma(N+1)}{\Gamma(N-b+k)}\left \langle \left(\sum_j n_j\right)^{-b} n_1^k \prod_{i=1}^{b-k}n_{i+1} \right\rangle 
\\& \approx \frac{\Gamma(N+1)}{\Gamma(N-b+k)\Gamma(b)}\int_0^\infty dz\, z^{b-1} \langle n^k e^{-z n}\rangle  \langle n e^{-zn}\rangle^{b-k}  \langle e^{-zn}\rangle^{N-b+k-1} \\
&=  \frac{\Gamma(N+1)}{\Gamma(N-b+k)\Gamma(b)}\int_0^\infty dz\, z^{b-1} (-1)^{k+1} [\partial_z^k \Phi(z,t)]  [\partial_z \Phi(z,t)]^{b-k}  e^{-(N-b+k-1)\Phi} \end{split}
\end{equation}
Using $\partial_t \Phi(z,t)=\frac{\sigma^2}{2Dt}z^{-1} \Phi(z,t)$, defining $\gamma = \frac{\sigma^2}{2Dt}$, and assuming $N\gg b$, we have
\begin{equation}
\begin{split}
\Lambda_{b,k} &=  \frac{\gamma^{b-k}\Gamma(N+1)}{\Gamma(N-b+k)\Gamma(b)}\int_0^\infty dz\, z^{k-1}(-1)^{k+1} [\partial_z^k \Phi(z,t)]\Phi(z,t)^{b-k}  e^{-(N-b+k-1)\Phi} \\
&\approx  -\prod_{i=0}^{k-1}(i-\gamma)\frac{\gamma^{b-k}}{\Gamma(b)}\int_0^\infty dz\, z^{-1} [N\Phi(z,t)]^{b-k+1}  e^{-N\Phi}
\end{split}
\end{equation}
Changing integration variables to $\phi = N\Phi(z,t)$, we find
\begin{equation}
\begin{split}
\Lambda_{b,k} &\approx -\prod_{i=0}^{k-1}(i-\gamma) \frac{\gamma^{b-k-1}}{\Gamma(b)}\int_0^\infty d\phi\, \phi^{b-k}  e^{-\phi}\\
&\approx - \prod_{i=0}^{k-1}(i-\gamma) \frac{\gamma^{b-k-1}\Gamma(b-k+1)}{\Gamma(b)} = \gamma^{b-k} \frac{\Gamma(k-\gamma)(b-k)!}{\Gamma(1-\gamma)(b-1)!}
\end{split}
\end{equation}
For $t=\frac{\sigma^2}{2D}+\delta t$ with $\delta t \ll \frac{\sigma^2}{2D}$, $\gamma\approx 1$ and we find 
\begin{equation}
\begin{split}
\Lambda_{b,k} & \approx \delta t\frac{2D}{\sigma^2}\frac{(k-2)!(b-k)!}{(b-1)!} = \delta t \lambda_{b,k}
\end{split}
\end{equation}
which are the general BSC rates $\lambda_{b,k}$.

We have assumed a steady distribution with a well-defined cutoff at $x_c$.
The actual high fitness end of the population fluctuates quite a bit, which would result in large population size fluctuations if the population size was not tightly controlled via a mean fitness that is catching up. This feed back via the mean fitness, however, is delayed and does not affect the dynamics at the nose instantaneously. The mapping to the FKPP models discussed below is better suited to describe the effects of fluctuations of the nose on coalescence.

\section{Relation to FKPP models}
After the equilibration time $T_{eq}$ the ancestors of all extant individuals
are located in a region of width $\propto D^{1/3}$ in the nose of the fitness
distribution roughly $x_c \approx \frac{\sigma^4}{4D}$ ahead of the mean. Within
this region, the growth rate does not vary much and the deterministic population
distribution is a solution of
\begin{equation}
D\frac{\partial c(x)}{\partial x^2} - \sigma^2 \frac{\partial c(x)}{\partial x} + x_c c(x) =0
\end{equation}
which has an exponential solution $c(x) \sim e^{-x \gamma}$ with $\sigma^2(\gamma) =
\frac{x_c}{\gamma}+D\gamma$. Waves of this form select the minimal velocity
corrected by a cut-off \cite{Brunet:1997p18870}, which corresponds to $\sigma^2 =
2\sqrt{x_c D}$ and $\gamma_0 = \sigma^2/2D$, consistent with \cite{Cohen:2005p45154}. 
Brunet et al.~\cite{Brunet:2006p47336,Brunet:2007p18866} have argued that the coalescence
process of such waves is of Bolthausen-Sznitman type with a typical time scale
given by
\begin{equation}
\label{eq:supp_FKPP_Tc}
T_c = \frac{\log^3 n}{\pi^2 \gamma_0^3 \alpha} = \frac{\log^3 n}{2\pi^2 x_c}
\end{equation}
where $\alpha = 2x_c \gamma_0^{-3}$ is the second derivative of $\sigma^2(\gamma)$ with respect to 
$\gamma$ at $\gamma=\gamma_0$ and $n$ is the size of the population relevant for the FKPP wave. In our case,
only the part of the wave that has a substantial chance of being the common
ancestor of the population is relevant and this part is located in an area of
with $\mathcal{O}(D^{1/3})$ away from the tip. The number of individuals in this
range is
\begin{equation}
\log n \propto \log \int_{-D^{1/3}}^0 dx\, e^{-\frac{x\sigma^2}{2D}} \propto \frac{\sigma^2}{D^{2/3}} 
\end{equation}
Substituting this into \EQ{supp_FKPP_Tc} and using $x_c = \sigma^4/4D$, we find
\begin{equation}
T_c \propto \frac{\sigma^2}{D}
\end{equation}
in agreement with the above findings. Furthermore, this argument supports our finding of
Bolthausen-Sznitman coalescence after the branching process description is no longer valid.

\section{The BSC in tuned models}
The calculation giving rise to Eq.~\EQbsrates~of the main text was done for a specific model of
selection and mutation. We now want to argue that similar
considerations hold  for a large class of related models.
The evolution and composition of traveling evolutionary waves is governed by the
fitness distribution in the population, $c(\x)$, and the probability that an
individual at fitness $\x$ will take over the entire population $u(\x)$
\cite{Hallatschek:2011p39697}. The deterministic and linear parts (i.e., up a cutoff) of $u(\x)$ and $c(\x)$
are left and right eigenfunctions of the same evolution operator $\OL$. 
The distribution $c(\x)$ is a localized pulse, while $u(\x)$ is an increasing
function of $\x$ (high fitness individuals have a higher chance of
spreading than low fitness individuals). 

Looking forward in time, $u(\x)$ is the probability that a lineage at $\x$ is the
common ancestor of the entire population in the distant future. Before the
descendants of a single lineage take over the population, they spread evenly
across the fitness distribution. Once spread out, every individual is equally likely to be the
common ancestor and the chance that one descendant of
this lineage fixes is given by the fraction $f(\x)$ of the total
population made up by descendants of the lineage. Hence $u(\x)$ is roughly the product of the 
probability that the lineage survives up to the time when it's descendants are equally distributed 
across the population (establishment), and the expected fraction $f(\x)$ conditional on establishment.
Since only large values of $\x\gg \sigma$ are relevant and the establishment probability in tuned
models is $p_{est}(\x)\sim \x/2$, we expect $f(\x) \approx 2u(\x)/\x$.

To analyze the coalescent properties, we need to know the distribution $p(f)$ of population fractions
that trace back to single individuals a time $T_{eq}$ in the past, where $T_{eq}$ is the
equilibration time of individuals across the population ($T_{eq}\approx \sigma^2/2D$ in our case). 
\begin{equation}
p(f) = c(\x)p_{est}(\x) \left(\frac{d f}{dx}\right)^{-1}
\end{equation}
If the evolution operator $\OL$ is dominated by selection all individuals that give rise to large $f$
come from the high-fitness tail. In this region, $f(\x)$ rises from very low values to $\mathcal{O}(1)$, 
while the product of $f(\x)c(x)$ (or equivalently $u(\x)c(x)$) is roughly constant (viewed as a function of $f$) \cite{Hallatschek:2011p39697}. 
Furthermore, $f(\x)$ grows exponentially with a power of $\x$, implying that $f'(\x) = g(\x)f(\x)$ where $\log g(\x)$ is a slowly
varying function of $\x$. Using $f'(\x)\sim f(\x)$ and $c(\x)\sim f(\x)^{-1}$, we find
\begin{equation}
p(f) \sim f^{-2}
\end{equation}
Given this effective off-spring distribution, we can calculate the coalescence probabilities
\begin{equation}
\lambda_{b,k} \sim \int df \, p(f) f^k(1-f)^{b-k} = \frac{(k-2!)(b-k)!}{(b-1)!}
\end{equation}
which are proportional to the BSC rates. 
The constant of proportionality depends on the details of the stochastic
dynamics which we have calculated above by other means (see also \cite{Hallatschek:2011p39697} 
for an alternative calculation of $Q_2(T_{eq}) = \int d\x c(\x)u^2(\x)$). 

\subsection*{Distribution of ancestors}
If $u(\x)$ is the probability that a lineage with log-fitness $\x$ survives to the present,
the log-fitness values of common ancestors of the population are distributed as $c(\x)u(\x)$. 
Now consider the MRCA common ancestor which lived a long time in the past. 
By definition, the MRCA gives rise to at least two branches that survive to the present. 
We therefore expect that the log-fitness of an MRCA that gives rise to $k$ surviving branches
should be approximately distributed as $c(\x)u^k(\x)$.

We measured the distribution of log-fitness of all common ancestors,
$\mathrm{CA}$, of the population and determined $u(\x) = \mathrm{CA}(\x)/c(\x)$,
see \figAncestors~of the main text. We then compared the distribution of the
log-fitness of the MRCAs of the population to $c(\x)u^2(\x)$, finding good
agreement with simulations. The data is too noisy to distinguish different $c(\x)u^k(\x)$
for different $k>2$.

\section{Laplace transform of pair coalescence times}
Instead of simulating individuals, one can also track the population density spread out
on a 2 dimensional grid. To characterize diversity in an adapting population, we designate 
one dimension as the log-fitness axis, while the dynamics along the
other direction is neutral direction. This keeps a record of the pair-coalescence time distribution 
as follows.
 
The distribution of pairwise distance between individuals along the neutral axis can be
decomposed into the distribution of pairwise coalescence times and the distribution of
distances along the neutral axis, given the coalescence time $T_2$. If individuals
hop left or right with probability $\mu/2$ each, the distance distribution along the neutral direction
follows a diffusion equation with diffusion constant $D=\mu/2$. Hence the
averaged distribution over all pairwise differences along the neutral coordinate
should be
\begin{equation}
\label{eq:appLaplace}
P(\Delta x)  = \int_0^\infty dT P(T) \frac{e^{-\frac{\Delta x^2}{2DT}}}{\sqrt{2\pi D T}}
\end{equation}
Fourier transforming this distribution in $\Delta x$ results in 
\begin{equation}
\hat{P}(k)  = \int_0^\infty dT P(T) e^{-\frac{ k^2 DT}{2}}
\end{equation}
Hence the Fourier transform of the pair-wise distance function is the Laplace
transform of the pair-coalescence time in the variable $k^2 D/2$.

We know from the individual based simulations that the coalescence process is
essentially broken into two phases, a waiting period during which nothing is happening,
followed by an exponential decay with charactistic time $T_c$. The Laplace transform of such a function is
\begin{equation}
\hat{P}(z) = \frac{e^{-T_{delay} z}}{1+T_{c}z}
\end{equation}
These two parameters can be easily extracted by fitting this function to simulation results. Note that this Laplace transform with $z=\mu$ equals the probability of identity by state of two randomly chosen
individuals.

\section{Simulations}
\subsection{Forward simulations of adapting populations}
In our stochastic simulations, each individual of the population is characterized by a log-fitness that determines the expected number of offspring of this individual in the following generation. The number of offspring is Poisson distributed with a mean $\exp(\xu-\lambda)$, where $\lambda = \mfit- (1-N/N_0)$ is a density regulating factor that keeps the population size approximately at $N_0$. This type of density regulation is indistinguishable from a Fisher-Wright model with exactly constant population size if $N_0$ is large, in which case the actual population size deviates from $N_0$ only by fluctuations of order $\sqrt{N_0}\ll N_0$. Individual fitness $\xu$ is updated via mutations drawn from a specified distribution $K(\delta)$ (in most cases a Gaussian, but we use other distributions too). The population is propagated asexually.

As the population evolves, we keep track of each individual's parent and construct a genealogy that links each  individual in the present generation to all its ancestors. All nodes of the genealogy that don't have any offspring in the present population are deleted. From the genealogy, properties like the pair coalescent times can be easily calculated by tracing the lineages of randomly chosen individuals until they meet. Similarly, the most recent common ancestor of a population sample or the entire population can be determined this way. 

Nodes in the genealogy also store how many individuals descent from them. This is convenient when calculating the site frequency spectra since a mutation that happens in a particular node of the genealogy is going to be present in all its descendants (ignoring back mutations). Since neutral mutations don't affect the population dynamics or the genealogies, we can calculate the neutral SFS simply by summing over all opportunities for neutral mutations in the genealogy.

\subsection{Bolthausen-Sznitman coalescent simulations}
The direct Bolthausen-Sznitman coalescent simulations follow a backward process starting with a population of $N$ individuals. The time of the next merger event is drawn from an exponential distribution with rate given by Eq.~8 in the main text. Then the type of event and the $k$ individuals that merge are chosen according to Eq.~7 of the main text. This process is repeated until all lineages have coalesced.

\subsection{Measuring the Laplace transform of $P(T_2)$ in large populations}
To measure the Laplace transform of the pair coalescence time distribution, we represent the population as a cloud on a two dimensional grid. This cloud spreads by diffusion in all directions, but is selected along one direction of grid. Diffusion is approximated as a discrete hopping process with different hopping rates in the selected and unselected directions. As explained above (\EQ{appLaplace}), the distribution of the cloud in the direction orthogonal to the direction of selection is a measure of the pair coalescence time. This distribution is measured and saved for further analysis.

\subsection{Code}
The source code of programs and scripts is available from the author's website: \\

\url{http://www.eb.tuebingen.mpg.de/research/research-groups/richard-neher/theoretical-population-genetics.html}

\clearpage
\begin{figure}[htp]
\begin{center}
  \includegraphics[width=0.7\columnwidth]{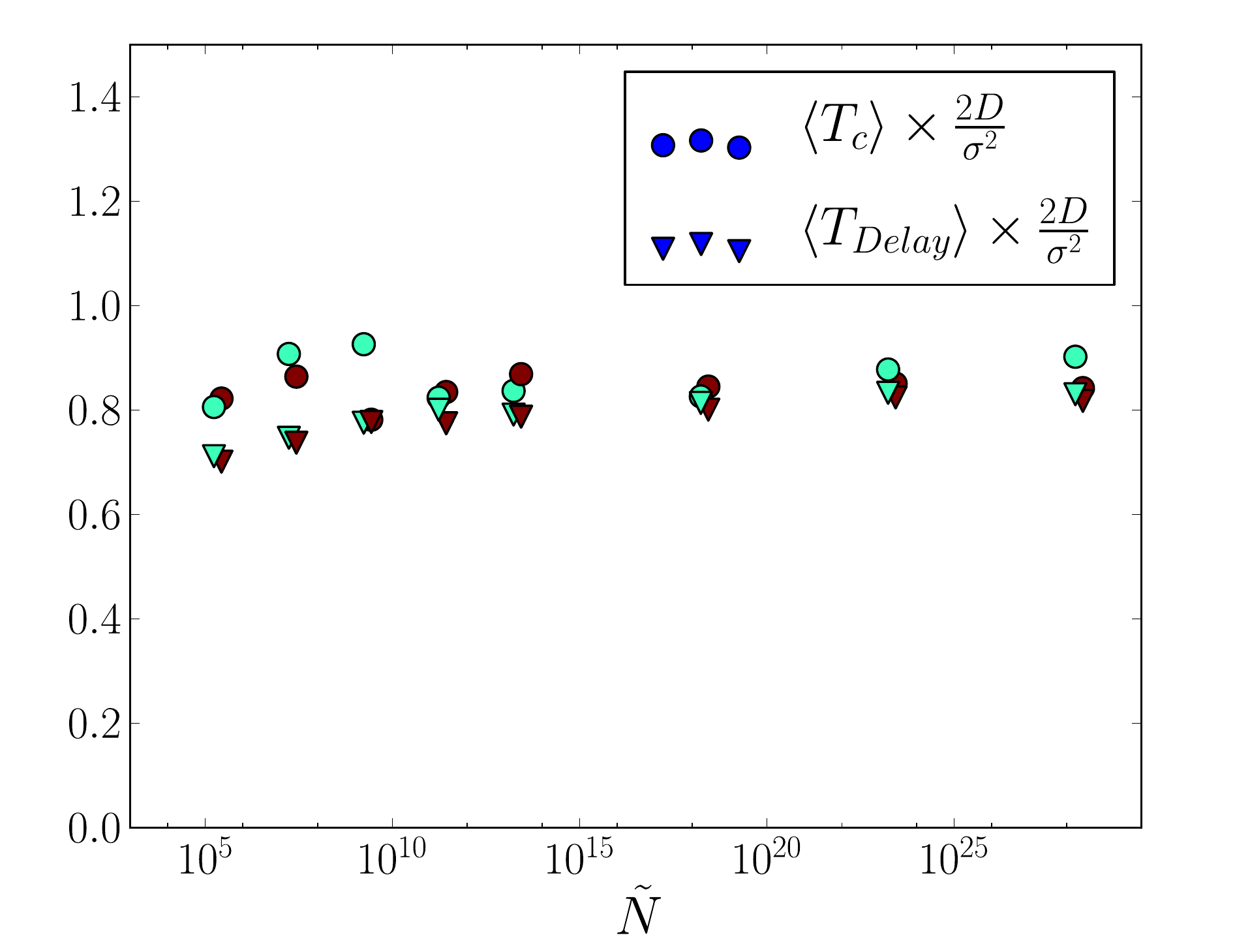}
  \caption[labelInTOC]{The delay $T_{delay}$ and the coalescence time scale $T_{c}$ can be extracted from the Laplace transform of $P(T_2)$ for very large populations and confirms that $T_c\approx \frac{\sigma^2}{2D}$.}
  \label{fig:applargeN}
\end{center}
\end{figure}

\begin{figure}[htbp]
\begin{center}
 \includegraphics[width=\columnwidth]{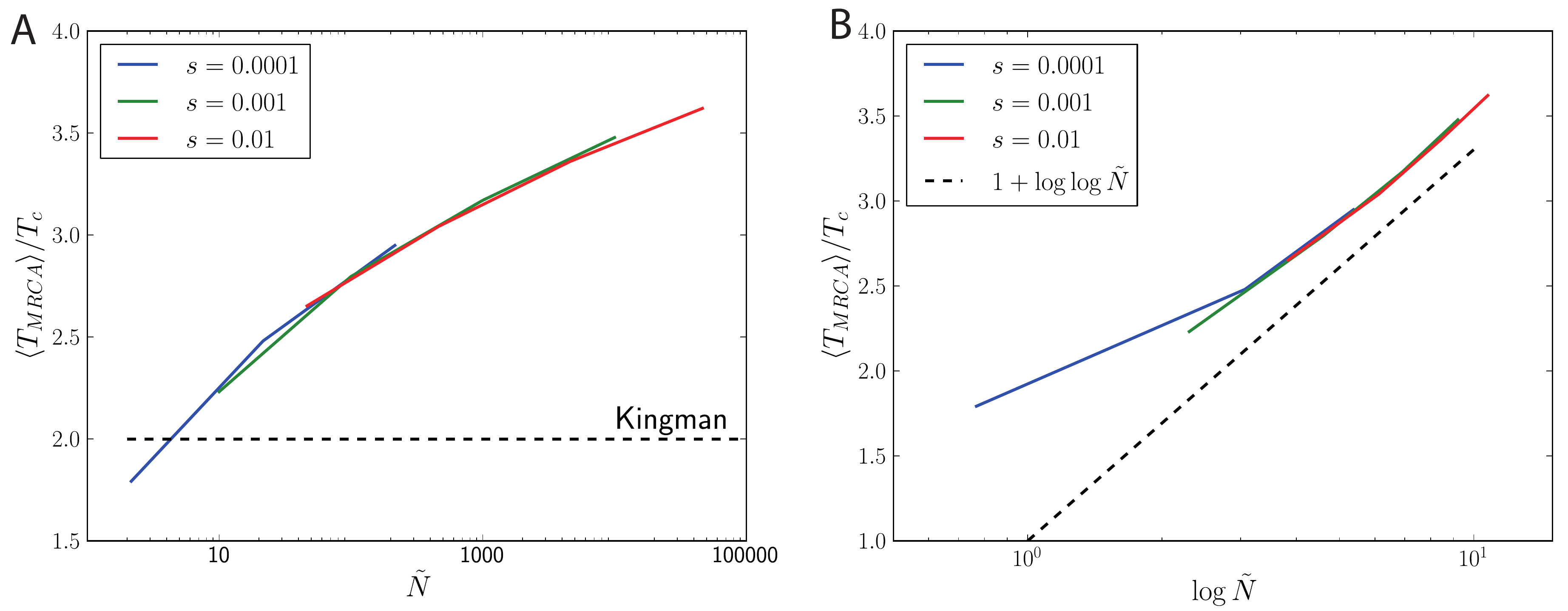}
 \caption{The average time to the most recent common ancestor of the
   whole population increases relative to the
 time scale of coalescence with the population size $\tilde{N}=ND^{1\over 3}$. This is a well known feature
 of the Bolthausen-Sznitman coalescent where one expects $\langle T_{MRCA} \rangle = T_c \log\log N$. Panel
 B compares $\langle T_{MRCA} \rangle/T_c$ to $1+\log\log\Ne$, where the additional $1$ is necessary to account
 for $T_{delay}\approx T_c$ before coalescence sets in.
}
 \label{fig:TMRCA}
\end{center}
\end{figure}

\begin{figure}[htbp]
\begin{center}
 \includegraphics[width=0.6\columnwidth]{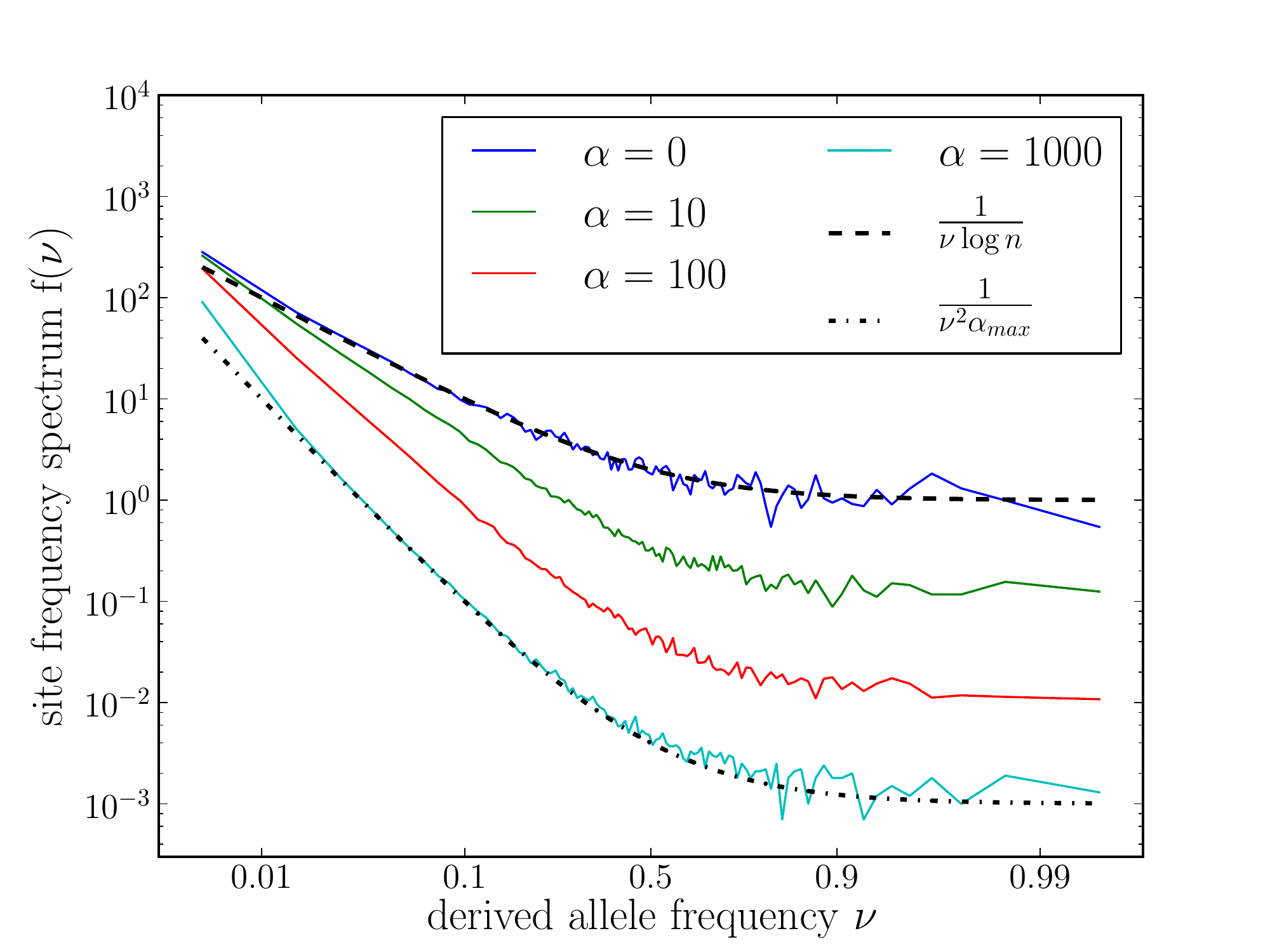}
 \caption{The site frequency spectrum of neutral derived mutations in Kingman's coalescent with constant ($\alpha=0$) and
 exponentially growing population sizes ($\alpha>0$ is the growth rate measured in units of $N^{-1}$). The genealogies are produced with the program \texttt{ms} \cite{Hudson:2002p37266}. The black lines show the theoretical expectation for a population of constant size and a rapidly expanding population. The $x$-axis is scaled as in Fig.~\affigure~of the main text.}
 \label{fig:af_exp_growing}
\end{center}
\end{figure}

\begin{figure}[htbp]
\begin{center}
  \includegraphics[width=\columnwidth]{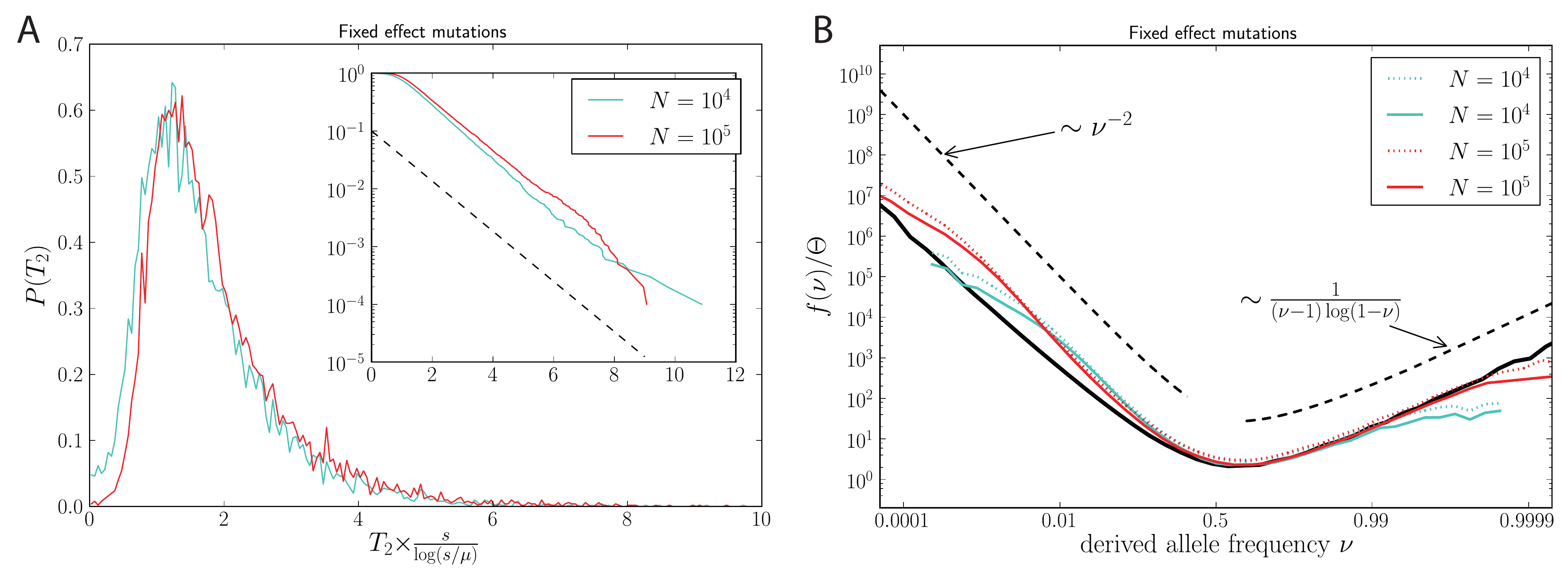}
  \caption[labelInTOC]{Results for an alternative models of adaptation, in which all
    mutations confer the same selective advantage $s$ and the mutation
    rate satisfies $\mu\ll s$. Panel A shows the distribution of pair
    coalescent times, panel B shows the site frequency spectra and the
    comparison to the Bolthausen-Sznitman coalescent. The times are rescaled with the prediction for the coalescent time $\langle T_2\rangle \approx  s^{-1}\log(s\mu^{-1})$ by Desai et al. \cite{desai_genetic_2012}. Solid lines correspond to $\mu/s = 0.01$, dotted lines to $\mu/s = 0.1$, while $s=0.01$. }
  \label{fig:supp_af_step}
\end{center}
\end{figure}
\begin{figure}[htbp]
\begin{center}
  \includegraphics[width=\columnwidth]{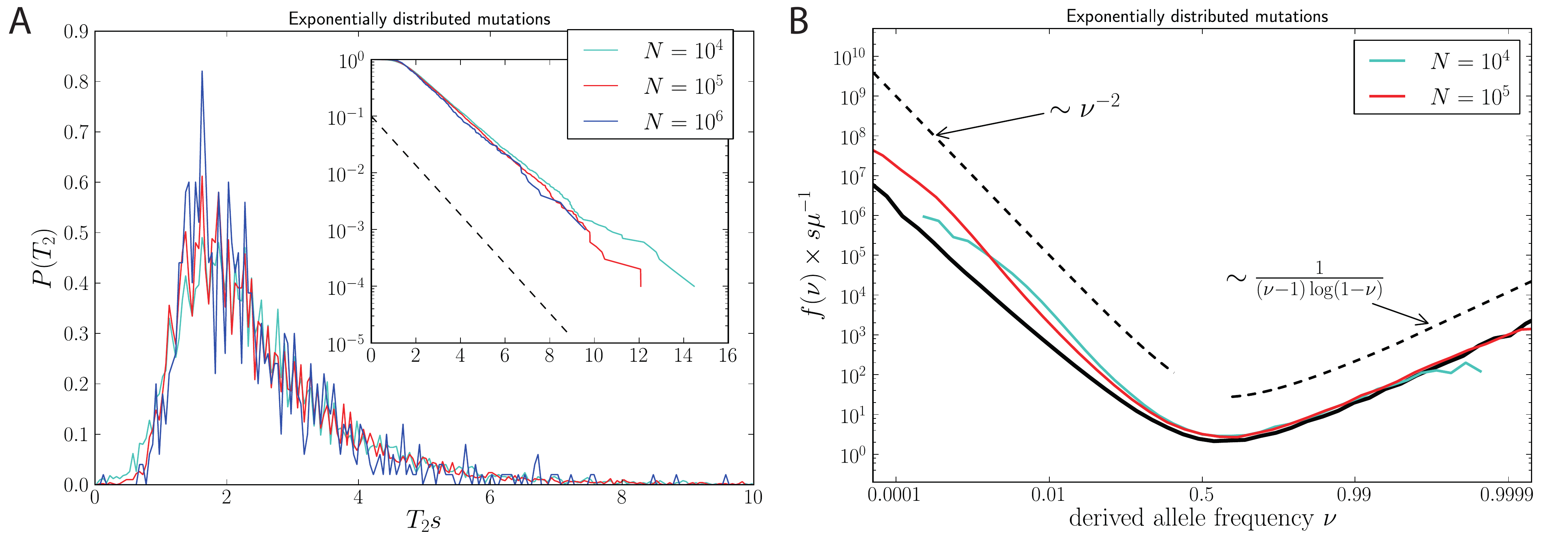}
  \caption[labelInTOC]{Results for an alternative model of adaptation, in which the
    selective advantage of new mutations is drawn from an exponential
    distribution with mean $s$, while the mutation rate satisfies $\mu\ll
    s$. Panel A shows the distribution of pair coalescent times, panel
    B shows the site frequency spectra and the comparison to the
    Bolthausen-Sznitman. In the absence of a prediction for the
    dependence of $T_2$ on $Ns$ or $\mu s^{-1}$, we rescale times by
    $s=0.01$. The mutation rate equals $\mu=0.1s$. }
  \label{fig:supp_af_exp}
\end{center}
\end{figure}
\begin{figure}[htbp]
\begin{center}
  \includegraphics[width=\columnwidth]{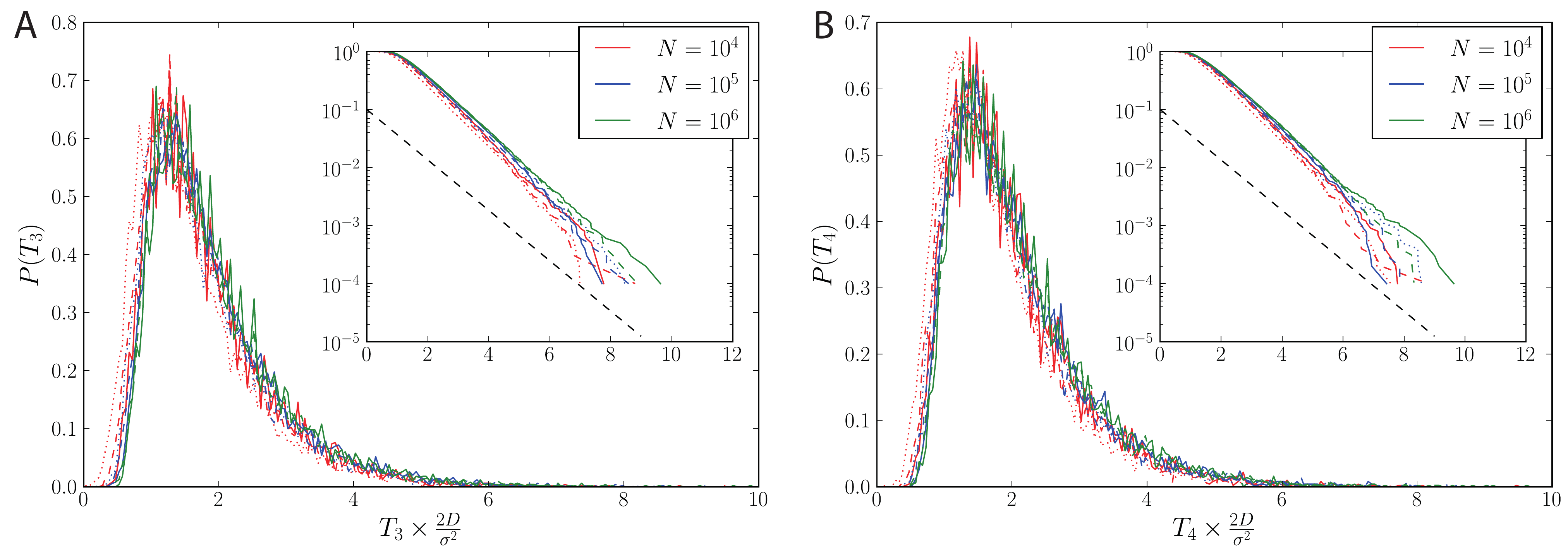}
  \caption[labelInTOC]{The distribution of the time to the most recent common
  ancestor of three individuals (panel A), and four individuals (panel B) in a model where mutational effects are normally distributed and mutations are frequent; compare Fig.~3 of the main text. Different line styles correspond to $s=0.01$ (solid),
  $s=0.001$ (dashed), and $s=0.0001$ (dotted), while the mutation rate is $\mu=1$. For each parameter combination, random pairs 
  are sampled at 10000 time points $2s^{-2/3}$ generations apart
    }
  \label{fig:supp_Tp}
\end{center}
\end{figure}

\end{document}